\documentclass[conference]{IEEEtran}

\usepackage{graphicx}
\usepackage{epstopdf}
\usepackage{balance}  % for  \balance command ON LAST PAGE  (only there!)
\usepackage{textcomp}

\makeatletter
\newif\if@restonecol
\makeatother

\usepackage[linesnumbered, ruled, vlined]{algorithm2e}

\usepackage{multirow}
\usepackage[caption=false]{subfig}
\usepackage{overpic}
\usepackage{color}

\usepackage{amsmath,amssymb,amsfonts,amsthm}
\usepackage{stmaryrd}

\usepackage{mathtools}

\newcommand{\eat}[1]{}
\newcommand{\M}[1]{\mathcal{#1}}
\newcommand{\SC}{sc_{max}}
\newcommand{\LB}{\mathcal{B}_\perp}
\newcommand{\TSPACE}{\vspace{-5pt}}
\newcommand{\EQSPACE}{\vspace{-5pt}}

\newtheorem{lemma}{Lemma}
\newtheorem*{definition}{Definition}
\newtheorem{condition}{Proposition}

\DeclareCaptionType{copyrightbox}

\begin{document}

\title{A Technical Report: Entity Extraction using Both Character-based and Token-based Similarity}

\author{%
% author names are typeset in 11pt, which is the default size in the author block
{Zeyi Wen{\small $^{\#}$}, Dong Deng{\small $^{*}$}, Rui Zhang{\small $^{\#}$}, Kotagiri Ramamohanarao{\small $^{\#}$} }%
% add some space between author names and affils
\vspace{1.6mm}\\
\fontsize{10}{10}\selectfont\itshape
% 20080211 CAUSAL PRODUCTIONS
% separate superscript on following line from affiliation using narrow space
$^{\#}$University of Melbourne, Victoria, Australia\\
\fontsize{9}{9}\selectfont\ttfamily\upshape
%
% 20080211 CAUSAL PRODUCTIONS
% in the following email addresses, separate the superscript from the email address 
% using a narrow space \,
% the reason is that Acrobat Reader has an option to auto-detect urls and email
% addresses, and make them 'hot'.  Without a narrow space, the superscript is included
% in the email address and corrupts it.
% Also, removed ~ from pre-superscript since it does not seem to serve any purpose
\{zeyi.wen, rui.zhang, kotagiri\}@unimelb.edu.au\\
%$^{3}$\,third.author@first-third.edu%
% add some space between email and affil
\vspace{1.2mm}\\
\fontsize{10}{10}\selectfont\rmfamily\itshape
% 20080211 CAUSAL PRODUCTIONS
% separated superscript on following line from affiliation using narrow space \,
$^{*}$Massachusetts Institute of Technology, USA\\
\fontsize{9}{9}\selectfont\ttfamily\upshape
% 20080211 CAUSAL PRODUCTIONS
% removed ~ from pre-superscript since it does not seem to serve any purpose
dongdeng@csail.mit.edu
}

\maketitle

\begin{abstract}
Entity extraction is fundamental to many text mining tasks such as organisation name recognition.
A popular approach to entity extraction is based on matching sub-string candidates in a document against a dictionary of entities.
To handle spelling errors and name variations of entities,
usually the matching is approximate and edit or Jaccard distance is used to measure dissimilarity between
sub-string candidates and the entities.
For approximate entity extraction from free text,
existing work considers solely character-based or solely token-based similarity
and hence cannot simultaneously deal with minor variations at token level and typos.
In this paper, we address this problem by considering both character-based similarity and token-based similarity (i.e. two-level similarity).
Measuring one-level (e.g. character-based) similarity is computationally expensive,
and measuring two-level similarity is dramatically more expensive.
By exploiting the properties of the two-level similarity and the weights of tokens,
we develop novel techniques to significantly reduce the number of sub-string candidates
that require computation of two-level similarity against the dictionary of entities.
A comprehensive experimental study on real world datasets show that our algorithm can efficiently extract entities from documents and produce
a high F$_1$ score in the range of [0.91, 0.97].
\end{abstract}

\section{Introduction}
In text mining, a primitive task is entity extraction---the recognition of the names of entities such as people,
locations and organisations---in a free text document.
A common approach to entity extraction is to compare sub-strings of a document
(hereafter ``\textit{sub-string candidates}'' or simply ``\textit{candidates}'')
against a dictionary of entities,
and the approach has wide use in applications such as named entity recognition (NER)~\cite{nadeau2007survey}.
This approach needs to handle the following two issues.
(i) Orthographical or typographical errors (\textit{typos} hereafter) may appear in documents.
For example, ``Oxford" may be incorrectly written as ``Oxfort''.
(ii) Different names may refer to the same entity.
For example, ``Oxfor\underline{t} University" and ``Univer\underline{c}ity of Oxford" are the same as ``The University of Oxford".
Addressing the two issues in the context of free text is very challenging,
since every word of the free text may be the starting (or ending) position of an entity in the dictionary.
As a result, the number of sub-string candidates is very large,
and all those sub-string candidates need to match against each entity in the dictionary.
Previous methods~\cite{wang2009efficient, deng2012efficient, chakrabarti2008efficient}
using only character-based or token-based (i.e. one-level) similarity cannot handle \textit{both} of the issues in the free text context.

To the best of our knowledge, no existing work has addressed entity extraction from free text using two-level similarity.
This is the first work to investigate entity extraction from free text using two-level similarity.
In this paper, we propose an algorithm by considering both character-based and token-based similarity
(i.e. two-level similarity) to extract entity from free text.
Measuring one-level similarity is computationally expensive,
and measuring two-level similarity is dramatically more expensive.
Without novel techniques to support the two-level similarity based algorithm,
extracting entities from a large number of documents against a large dictionary is computationally very expensive.
We observe that  a sub-string candidate can be similar to an entity only if they share some tokens, 
thus we first identify all the matched tokens from the document for each entity. 
Then based on the matched tokens, we enumerate all the sub-string candidates that potentially similar to the entity.
By exploiting the properties of the two-level similarity and the weights (measured by IDF~\cite{manning2008introduction}) of tokens,
we further develop a spanning-based method to dramatically reduce the number of sub-string candidates that require computation of two-level similarity.
To summarise, we make the following key contributions.
\begin{itemize}
	\item This is the first work to address the problem of approximate entity extraction from free text
	      using both character-based and token-based similarity.
		  We formulate the problem and propose novel techniques to solve the problem.
    \item For each entity, by naively enumerating $k$ matched tokens in a document to an entity,
          the total number of sub-string candidates produced is about $k^2$.
          By avoiding enumerating very short or very long sub-string candidates,
          we design a technique to reduce the number of sub-string candidates to
          $(u-l) \cdot k$, where $[l,u]$ is the range of the number of tokens in the candidates
          that are neither too short nor too long.
	\item By exploiting the properties of the two-level similarity and the weights of tokens,
		  we develop a spanning-based candidate producing technique to significantly reduce the number of sub-string candidates
		  to just $k$.
		  The key novelty of our spanning-based candidate producing technique lies in a novel lower bound and computation reuse strategy.
	\item We conduct extensive experiments to validate the efficiency and effectiveness of our algorithm.
		  The experimental results show that our algorithm can efficiently extract entities from documents,
		   and produce a high F$_1$ score in the range of [0.91, 0.97].
\end{itemize}

The rest of the paper is organised as follows.
In Section~\ref{paper:rw}, we discuss related work in entity extraction.
Then, we present preliminaries in Section~\ref{paper:pre},
and describe our algorithm using a two-level similarity in Section~\ref{paper:2ED}.
In Section~\ref{paper:es}, we report experimental results of our algorithm.
Finally, we conclude the paper in Section~\ref{paper:con}.

\section{Related Work}
\label{paper:rw}
We categorise the related work of entity extraction into two main groups:
studies based on machine learning approaches and studies based on string matching approaches.
Our work falls into the later group.

\textbf{Machine learning based approaches}:
Carreras et al. proposed an Adaboost based approach for named entity extraction~\cite{carreras2002named}.
Their key idea is to extract entities using two classifiers:
a local classifier for detecting if a token belongs to a named entity;
a global classifier for detecting if a sub-string candidate is a named entity.
Jain and Pennacchiotti~\cite{jain2010open} proposed an approach using heuristics (e.g. tokens with first letter capitalised)
to extract entities from query log, and then the extracted entities are grouped into different clusters and assigned labels accordingly.
Cohen and Sarawagi~\cite{cohen2004exploiting} designed an algorithm using the Markov model for entity extraction.
The algorithm has two main phases.
First, a label (e.g. person name) is assigned to each token based on dictionaries/heuristics.
Second, the Markov model is trained and used to predict the entity probability for each sub-string candidate based on the token labels.
One major limitation of the abovementioned approaches is that they require significant amount of human effort to collect training datasets
and/or to tune heuristics.

\textbf{String matching based approaches}:
The approximate entity extraction problem can be viewed as the approximate string matching problem which is a well-studied problem.
Navarro gives a nice survey for the approximate string matching problem~\cite{navarro2001A}.
Here, we focus on some recent work in entity extraction.

Gattani et al. developed a dictionary-based algorithm for entity extraction~\cite{gattani2013entity}.
But their algorithm aims to extract sub-string candidates that exactly match entities in dictionary from short documents (e.g. tweets).
Kim et al.~\cite{kim2005n} proposed a memory efficient indexing approach for string matching using character-based similarity.
Their proposed index is memory friendly by reusing position information of $n$-grams through a two-level scheme.
Wang et al. proposed an approximate entity extraction algorithm using neighbourhood generation~\cite{wang2009efficient}.
Deng et al.~\cite{deng2012efficient} designed an efficient algorithm for approximate entity extraction based on trie tree index.
Kim and Shim proposed an algorithm that finds from a document top-$k$ most similar sub-string candidates to an entity~\cite{kim2013efficient}.
A more recent study~\cite{wang2016local} presents techniques to find duplicated text segments between two documents using token-level similarity.
All these algorithms use one-level, i.e. character-based or token-based, similarity to find similar entities (or text segments) in documents.

Some existing studies~\cite{hadjieleftheriou2010weighted, chaudhuri2003robust} designed similarity functions
and indexing techniques for the string similarity search problem.
Cohen et al.~\cite{cohen2003comparison} developed an open source software toolkit, which supports different similarity functions,
for measuring the similarity between two strings.
Chakrabariti et al.~\cite{chakrabarti2008efficient} proposed a filter using the token-based similarity to classify
sub-string candidates into two classes: valid sub-string candidates that may match some entities in the dictionary;
invalid sub-string candidates that do not match any entities in the dictionary.
The above work differs from ours, since we aim at developing an effective and efficient algorithm to extract entities from free text
using both character-based and token-based similarity.

\section{Preliminaries}
\label{paper:pre}

For ease of presentation, a token (e.g. word) of a sub-string candidate is called a \textit{text token}.
Similarly, we call a token of an entity in the dictionary an \textit{entity token}.
Some frequently used symbols in the rest of the paper are summarised in Table~\ref{tbl:symbols}.
In this section, we first give an approach to computing the weights (i.e. importance) of entity tokens and text tokens.
Second, we provide background knowledge of edit and Jaccard similarity.
Then, we present an algorithm for finding from a document text tokens which match any token in the dictionary.
Lastly, we formally define the approximate entity extraction problem.

\begin{table}
\centering
\begin{small}
\caption{Frequently used symbols}
\begin{tabular}{|c|l|} \hline
$t$, $e$, $s$						& a token, an entity token and a text token\\\hline
$idf(t)$, $w(t)$						& IDF and the weight of $t$, respectively \\\hline
$\mathcal{E}$, $\mathcal{S}$, $\mathcal{E}_i$,	$\mathcal{S}_j$	& an entity, sub-string candidate, the $i^{th}$ token	\\
								&  of $\mathcal{E}$, and the $j^{th}$ token of $\mathcal{S}$, respectively 	\\\hline
$eds(e$, $s)$		& the edit similarity of $e$ and $s$ \\\hline
$\tau, \delta$							 	& token and entity edit similarity thresholds \\\hline
%$\mathcal{C}$						& a set of core tokens of $\mathcal{E}$	\\\hline
%$\mathcal{C}_i$					& the $i^{th}$ core token in $\mathcal{C}$	\\\hline
%$E$								& a dictionary of entities	\\\hline
\end{tabular}
\end{small}
\label{tbl:symbols}
\TSPACE
\end{table}

\subsection{Assigning weights to tokens}
\label{paper:weighting}
In many applications,
the tokens of an entity (or a sub-string candidate) have different importance,
called \textit{weights} hereafter, in the entity (or the sub-string candidate).
Following common practice, we use IDF~\cite{manning2008introduction} to measure the weights of tokens.
In the approximate entity extraction problem, the dictionary is known a priori
and documents are unknown beforehand.
Hence, we compute the IDF value of a token based on the dictionary.
Specifically, given the dictionary with $N$ entities and a token $t$,
we count the number (denoted by $N_t$) of entities that contain $t$
to serve as the ``document frequency'' of the token.
Then, the IDF value of $t$ is computed by the following equation.
\begin{small}

\vspace{-6pt}
\begin{equation*}
idf(t) =  \log \frac{N}{N_{t} + 1}
%\label{eq:idf}
\end{equation*}
\end{small}%
The total IDF value of a set of tokens $\mathcal{A}$ is the sum of the IDF values of all the tokens in $\mathcal{A}$,
and can be computed as follows.
\begin{small}
\begin{equation}
T_{idf}(\mathcal{A}) = \sum_{t \in \mathcal{A}}{idf(t)}
\label{eq:tidf}
\end{equation}
\end{small}%
After computing $idf(t)$ and $T_{idf}(\mathcal{A})$,
we can compute the weight of the token $t$ by the equation below.
\begin{small}
\EQSPACE
\begin{equation}
w(t) = \frac{idf(t)}{T_{idf}(\mathcal{A})}
\label{eq:normalise}
\EQSPACE
\end{equation}
\end{small}%
Note that $\mathcal{A}$ can be either the entity $\mathcal{E}$ or the sub-string candidate $\mathcal{S}$.
We define the total weight of a subset $\mathcal{A}'$ of tokens in $\mathcal{A}$
(i.e. $\M{A}' \subseteq \M{A}$) as follows.
\begin{small}
\EQSPACE
\begin{equation}
T_w(\mathcal{A}') = \sum_{t \in \mathcal{A}'}{w(t)}
\label{eq:total_weight}
\EQSPACE
\end{equation}
\end{small}%

\subsection{Edit and Jaccard similarity}
\label{paper:char-ed}

\subsubsection{Edit similarity}
Edit-distance quantifies the dissimilarity of two tokens by counting the minimum number of edit operations
(i.e. deletion, insertion and substitution)
to transform from one token to the other.
Without loss of generality, we assume that all the edit operations have the same cost.

Based on edit-distance, edit similarity is to quantify the similarity of two tokens.
Formally, given two tokens $e$ and $s$, the edit similarity $eds(e, s)$ is defined as follows.
\begin{small}
\EQSPACE
\begin{equation}
eds(e, s) = 1 - \frac{ed(e, s)}{\max\{|e|, |s|\}}
\EQSPACE
\end{equation}
\end{small}%
where $ed(e, s)$ is the edit-distance between the two tokens;
$|e|$ and $|s|$ are the number of characters in $e$ and $s$, respectively.

\subsubsection{Jaccard similarity}
Jaccard similarity is mainly used as a token-based similarity.
In entity extraction, Jaccard similarity is for measuring the similarity
between an entity $\mathcal{E}$ and a sub-string candidate $\mathcal{S}$
and is defined as follows.
\begin{small}
\EQSPACE
\begin{equation}
JAC = \frac{|\mathcal{E} \cap \mathcal{S}|}{|\mathcal{E} \cup \mathcal{S}|}
	= \frac{|\mathcal{E} \cap \mathcal{S}|}{|\mathcal{E}| + |\mathcal{S}| - |\mathcal{E} \cap \mathcal{S}|}
\label{eq:jac}
\EQSPACE
\end{equation}
\end{small}%
where $|\mathcal{E} \cap \mathcal{S}|$ is the number of matched tokens between $\mathcal{E}$ and $\mathcal{S}$;
$|\mathcal{E} \cup \mathcal{S}|$ is the number of tokens in the union of $\mathcal{E}$ and $\mathcal{S}$;
$|\mathcal{E}|$ and $|\mathcal{S}|$ are the number of tokens in $\mathcal{E}$ and $\mathcal{S}$, respectively.

Note that the above-mentioned edit similarity is for character-based similarity,
while Jaccard similarity is for token-based similarity.
We postpone our definition of two-level similarity using edit or Jaccard similarity until Section~\ref{paper:2ED}.

\subsection{Matching text tokens against entities}
\label{paper:matching}

\begin{figure}
\center
\begin{overpic}[width=2.0in, height=1.2in]
{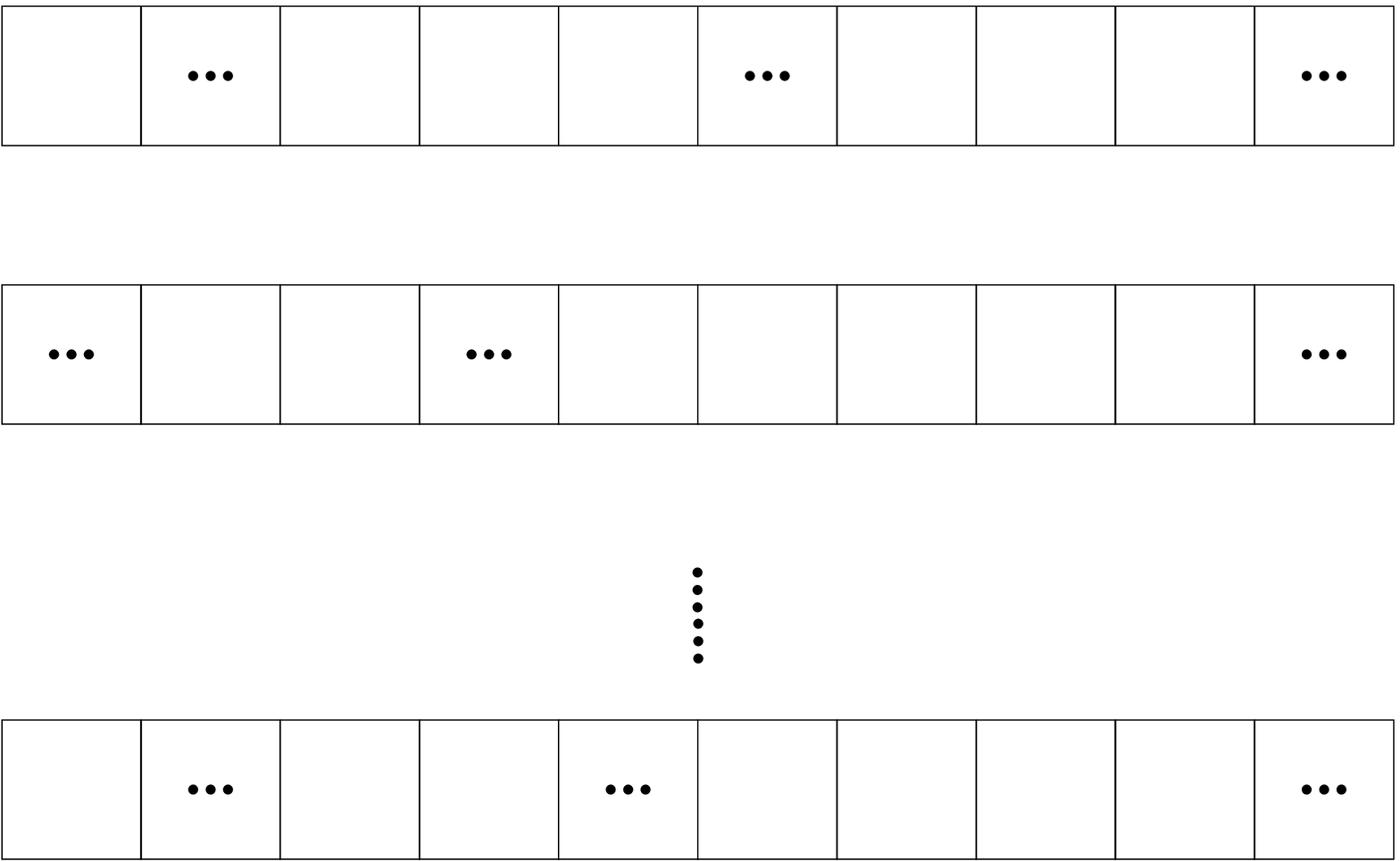}

\put(2, 52.5){$\mathcal{E}^1_{3}$}
\put(22, 52.5){$\mathcal{E}^1_{1}$}
\put(32, 52.5){$\mathcal{E}^1_{2}$}
\put(42, 52.5){$\mathcal{E}^1_{1}$}
\put(62, 52.5){$\mathcal{E}^1_{2}$}
\put(72.5, 52.5){X}
\put(81, 52.5){$\mathcal{E}^1_{3}$}
\put(4, 44){all the matched text tokens to $\mathcal{E}^1$}

\put(12, 33.5){$\mathcal{E}^2_{3}$}
\put(22, 33.5){$\mathcal{E}^2_{1}$}
\put(42, 33.5){$\mathcal{E}^2_{2}$}
\put(52, 33.5){$\mathcal{E}^2_{3}$}
\put(62, 33.5){$\mathcal{E}^2_{4}$}
\put(72, 33.5){$\mathcal{E}^2_{5}$}
\put(83, 33){X}
\put(4, 24){all the matched text tokens to $\mathcal{E}^2$}

\put(1, 3){$\mathcal{E}^N_{3}$}
\put(21, 3){$\mathcal{E}^N_{8}$}
\put(31, 3){$\mathcal{E}^N_{2}$}
\put(51, 3){$\mathcal{E}^N_{3}$}
\put(61, 3){$\mathcal{E}^N_{4}$}
\put(70.5, 3){$\mathcal{E}^N_{10}$}
\put(82.5, 3){X}
\put(3, -6){all the matched text tokens to $\mathcal{E}^N$}

\end{overpic}
\caption{Matched position of entities in a document}
\label{fig:matched-entity-doc}
\end{figure}

Since we are interested in extracting entities from documents (i.e. free text),
the first step is to find in the documents all the tokens that match to tokens in each entity of the dictionary.
We use  Li et al.'s algorithm~\cite{li2011faerie} for finding all the matched tokens in a document to an entity.
The details about how Li et al.'s algorithm works are unimportant for understanding our proposed algorithms.
Here, we briefly explain the results produced by the algorithm.
Figure~\ref{fig:matched-entity-doc} gives example results of the matched tokens in a document.
In the example, the dictionary contains $N$ entities which are denoted by $\mathcal{E}^{1}$, $\mathcal{E}^{2}$, ..., $\mathcal{E}^{N}$.
The rows represent the results of the same document matching to the $N$ entities.
A rectangle containing ``X'' indicates that the position\footnote{For ease of presentation,
we refer ``the position'' to ``the token at the position of the document''.} does not match any token of the entity;
a rectangle containing $\mathcal{E}^i_{j}$ indicates that the token at this position matches the $j^{th}$ token of the $i^{th}$ entity.

Example: Given a document $\mathcal{D}$ = ``\textit{... The Univercity of Oxfort is near the Oxford city ...}"
and an entity $\mathcal{E}^1 = $ ``The University of Oxford",
then each token of $\mathcal{E}^1$ is $\mathcal{E}^1_1 = $``The", $\mathcal{E}^1_2 = $ ``University", $\mathcal{E}^1_3 = $ ``of", and $\mathcal{E}^1_4 = $``Oxford".
After identifying all the matched token of $\mathcal{E}^1$ in $\mathcal{D}$,
we can represent $\mathcal{D}$ as ``... $\mathcal{E}^1_1$ $\mathcal{E}^1_2$ $\mathcal{E}^1_3$ $\mathcal{E}^1_4$ X X $\mathcal{E}^1_1$ $\mathcal{E}^1_4$ X ..."
(similar to Figure~\ref{fig:matched-entity-doc}).

\subsection{Problem definition}
\label{paper:pd}
We define our approximate entity extraction problem.

\begin{definition}[Approximate Entity Extraction]
Given an entity dictionary $E$, a document $\mathcal{D}$ and an entity similarity threshold $\delta$,
the approximate entity extraction problem is to find all pairs of entities and sub-string candidates
with similarity score not smaller than $\delta$.
I.e. $Sim(\mathcal{E}, \mathcal{S}) \geq \delta$
where $\mathcal{E}$ is an entity in $E$ and $\mathcal{S}$ is a sub-string candidate in $\mathcal{D}$.
\end{definition}

The similarity function $Sim(\mathcal{E}, \mathcal{S})$ (e.g. Fuzzy Jaccard) takes
both the character-based similarity and the token-based similarity into account
to measure the overall similarity of $\mathcal{E}$ and $\mathcal{S}$.
The weight of each token in $\mathcal{E}$ and $\mathcal{S}$ is measured by a weight function $w(\cdot)$.
An entity token $e$ and a text token $s$ are called ``similar'' or ``matched'' if the character-based edit similarity
of them is not smaller than $\tau$, i.e. $eds(e, s) \geq \tau$.

\section{Our proposed algorithm}
\label{paper:2ED}

\begin{figure}
\center
\begin{overpic}[width=3.1in, height=1.5in]
{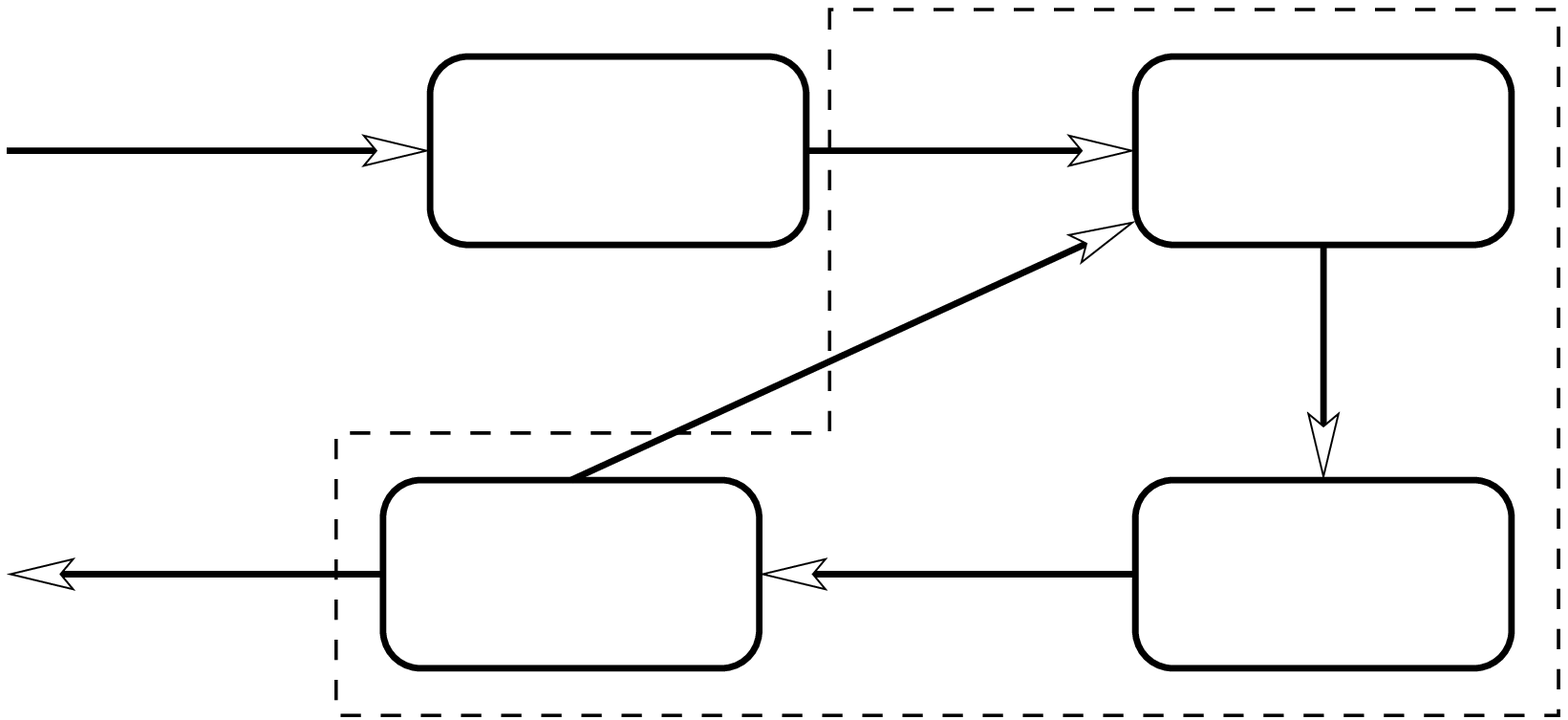}

\put(3, 40.5){documents}
\put(3.5, 35){entities in}
\put(0, 31.5){the dictionary}

\put(31.5, 39.5){Matching}
\put(28.5, 34.5){Text Tokens}

\put(51.5, 43.5){all matched}
\put(55.5, 40){tokens}
\put(52, 34.5){first entity}

\put(75.5, 39.5){Producing}
\put(74.5, 34.5){Candidates}

%\put(80, 16.5){\rotatebox{90}{candidates}}

\put(77, 11){Filtering}
\put(74.5, 6){Candidates}

\put(53, 11){candidates}
\put(54, 6){the entity}

\put(49, 19.5){\rotatebox{24}{next entity}}

\put(27.5, 11){Measuring}
\put(28, 6){Similarity}

\put(6.5, 14.5){extracted}
\put(8, 11){entities}

\end{overpic}
\vspace{-5pt}
\caption{Our algorithm for extracting entities}
\label{fig:framework}
\end{figure}

In this section, we elaborate our algorithm for entity extraction.
Figure~\ref{fig:framework} gives an overview of our algorithm
which has four components.
As matching text tokens in a document against a dictionary is well-studied,
the matching algorithm discussed in Section~\ref{paper:matching} serves in the Matching Text Tokens component
and proposing techniques to improve this component is out of the scope of this paper.
We focus on designing techniques for the other three components
which are inside the dashed line polygon in Figure~\ref{fig:framework}.
Our algorithm repeats the following three key steps until all the entities in the dictionary are checked.
Step (i): Based on the matched tokens output by Matching Text Tokens and a given entity in dictionary,
the Producing Candidates component produces all the sub-string candidates that may match to the entity.
Step (ii): The Filtering Candidates component filters out sub-string candidates that will not match to the entity.
Step (iii): The Measuring Similarity component computes the similarity
between each remaining sub-string candidate and the entity,
and outputs sub-string candidates with high similarity score as extracted entities;
then our algorithm goes back to Step (i) if any entity in the dictionary requires being checked.

Our algorithm can work with various similarity functions,
such as Jaccard similarity, Dice similarity, cosine similarity and edit similarity.
In this paper, we focus on designing techniques to our algorithm
using two-level edit-similarity (hereafter \textbf{FuzzyED})
and two-level Jaccard similarity (hereafter \textbf{Fuzzy Jaccard}),
as edit distance and Jaccard distance are two most commonly used distances in string matching.
Next, we first define the FuzzyED and Fuzzy Jaccard similarity.
Then, we present two algorithms to find sub-string candidates.
Finally, we provide filtering techniques to our algorithm for the corresponding two-level similarity.

\subsection{Two fuzzy similarity functions}

\subsubsection{The FuzzyED similarity}

Here, we define the cost of edit operations for token-based similarity and FuzzyED similarity.

\subsubsection*{Cost of edit operations on tokens of a candidate}
\label{paper:2ED-cost}
FuzzyED requires performing two levels of edit-distance:
the character-based edit-distance (for measuring similarity between two tokens) and
the token-based edit-distance (for measuring the similarity between an entity and a sub-string candidate).
As we have discussed the character-based edit-distance in Section~\ref{paper:char-ed},
here we provide details of the token-based edit-distance.

The total cost of FuzzyED is the cost of transforming a sub-string candidate to an entity.
Without loss of generality, we assume that only the sub-string candidate can be edited and the entity is not permitted to be edited.
We formulate the total cost of transforming a sub-string candidate $\mathcal{S}$
to an entity $\mathcal{E}$ using the following equation.
\begin{small}
\EQSPACE
\begin{equation}
FED(\mathcal{E}, \mathcal{S}) =  C_D(\mathcal{S}) + C_I(\mathcal{S}) + C_S(\mathcal{E}, \mathcal{S})
\label{eq:2ED-cost}
\EQSPACE
\end{equation}
\end{small}%
where $C_D(\mathcal{S})$ is the total deletion cost of removing text tokens from $\mathcal{S}$;
$C_I(\mathcal{S})$ is the total insertion cost of inserting entity tokens of $\mathcal{E}$ to $\mathcal{S}$;
$C_S(\mathcal{E}, \mathcal{S})$ is the total substitution cost of $\mathcal{E}$ and $\mathcal{S}$.
We let $\mathcal{S}'$ be a subset of tokens in $\mathcal{S}$ that match to $\mathcal{E}$;
$\mathcal{E}'$ denotes tokens that are matched by $\mathcal{S}'$.

\textit{Deletion}: The total deletion cost is computed by the following equation.
\begin{small}
\EQSPACE
\begin{equation*}
C_D(\mathcal{S}) = T_w(\mathcal{S} \setminus \mathcal{S}')
\EQSPACE
\end{equation*}
\end{small}%
where $\mathcal{S} \setminus \mathcal{S}'$ is a subset of the tokens in $\mathcal{S}$
(i.e. $\mathcal{S} \setminus \mathcal{S}' \subseteq \mathcal{S}$) that needs to be deleted from $\mathcal{S}$.

\textit{Insertion}: The total insertion cost is computed by the following equation.
\begin{small}
\EQSPACE
\begin{equation*}
C_I(\mathcal{S}) = T_w(\mathcal{E} \setminus \mathcal{E}')
\EQSPACE
\end{equation*}
\end{small}%
where $\mathcal{E} \setminus \mathcal{E}'$ is a subset of tokens in $\mathcal{E}$
(i.e. $\mathcal{E} \setminus \mathcal{E}' \subseteq \mathcal{E}$) that needs to be inserted to $\mathcal{S}$.

\textit{Substitution}: The total substitution cost is computed using the following equation.
\begin{small}
\EQSPACE
\begin{equation}
C_S(\mathcal{E}, \mathcal{S})= \sum_{e \in \mathcal{E}', s \in \mathcal{S}'}
{(1 - eds(e, s)) \times (w(e) + w(s))}
\label{eq:substitution}
\EQSPACE
\end{equation}
\end{small}%
where $s$ is a text token that matches the entity token $e$.
Next, we give the FuzzyED similarity based on the cost defined in Equation~\eqref{eq:2ED-cost}.

\subsubsection*{Computing the FuzzyED similarity}
\label{paper:2ED-similarity}
Given a sub-string candidate and an entity,
we can compute the total edit cost on transforming the sub-string to the entity by Equation~\eqref{eq:2ED-cost}.
We adapt the dynamic programming based algorithm~\cite{navarro2001A}
to compute the cost of the longest sub-string of $\mathcal{S}$
that is the most similar to the entity $\mathcal{E}$.
The time complexity of the dynamic programming based algorithm is $\mathcal{O}(mn)$,
where $m$ and $n$ are the number of tokens of $\M{E}$ and $\M{S}$, respectively.
The key idea of the dynamic programming based algorithm is similar to the algorithm for computing the character-based edit-distance
between two tokens.
We do not provide the details of the algorithm here
and suggest the interested readers to consult the original paper~\cite{navarro2001A}.

After computing the total edit cost in Equation~\eqref{eq:2ED-cost},
we can compute the FuzzyED similarity, denoted by $FEDS(\mathcal{E}, \mathcal{S})$, using the following equation.
\begin{small}
\vspace{-4pt}
\begin{equation}
FEDS(\mathcal{E}, \mathcal{S}) = \begin{cases}
											0	& \hspace{-0pt} \text{if $FED$}(\mathcal{E}, \mathcal{S}) > 1,\\
											1 - FED(\mathcal{E}, \mathcal{S}) & \hspace{32pt} \text{otherwise.}
										\end{cases}
\label{eq:2ED-sim}
\vspace{-4pt}
\end{equation}
\end{small}%
Note that the substitution cost may be larger than 1 when $\tau < 0.5$ (cf. Equation~\eqref{eq:substitution})
which results in {\small$FED(\M{E}, \M{S}) > 1$}.

A sub-string candidate $\mathcal{S}$ and an entity $\mathcal{E}$ are called ``matched'' or ``similar''
if {\small $FEDS(\mathcal{E}, \mathcal{S}) \geq \delta$}.

\subsubsection{The Fuzzy Jaccard similarity}
\label{paper:fj}

\begin{figure}
\center
\begin{overpic}[width=2.1in, height=0.7in]
{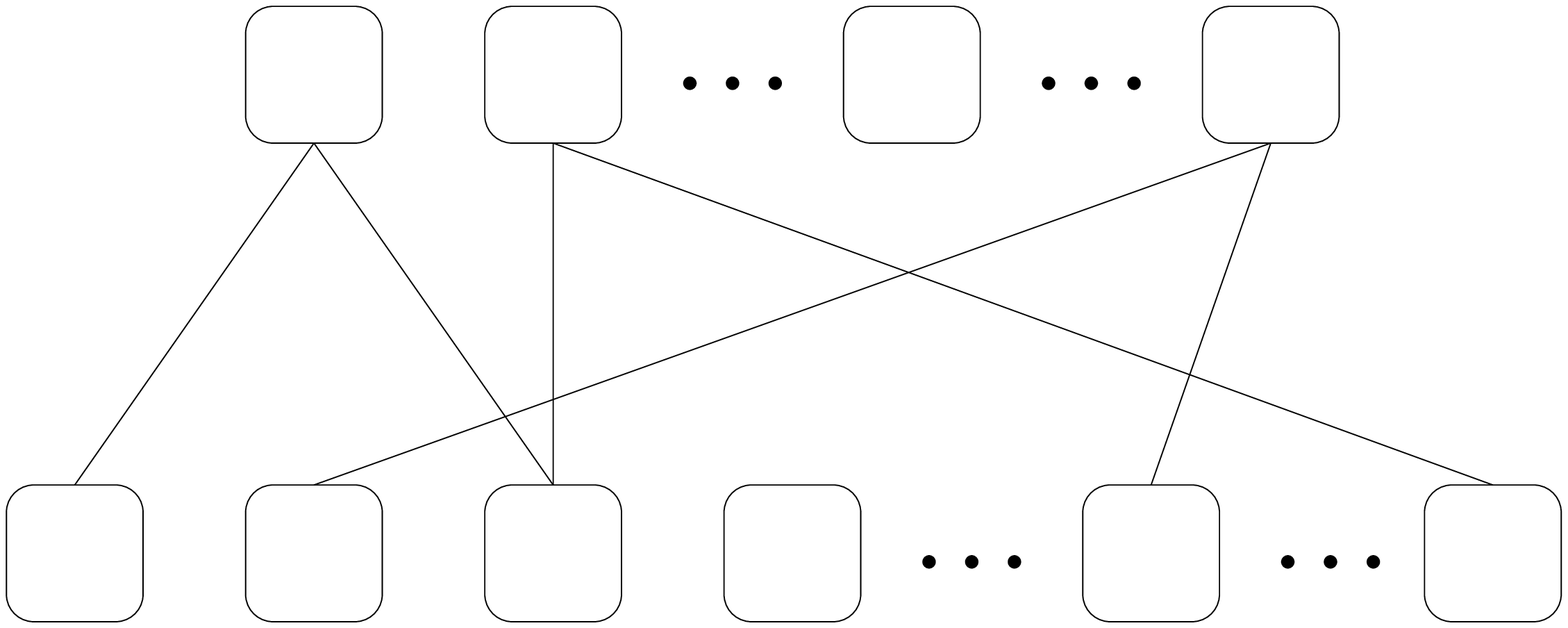}

\put(-8, 27.8){$\boldsymbol{\mathcal{E}}$}
\put(-8, 2.2){$\boldsymbol{\mathcal{S}}$}

\put(17.5, 27.8){$\mathcal{E}_1$}
\put(32.5, 27.8){$\mathcal{E}_2$}
\put(55.3, 27.8){$\mathcal{E}_i$}
\put(77.5, 27.8){$\mathcal{E}_m$}

\put(2, 2.2){$\mathcal{S}_1$}
\put(17, 2.2){$\mathcal{S}_2$}
\put(32, 2.2){$\mathcal{S}_3$}
\put(47.5, 2.2){$\mathcal{S}_4$}
\put(71, 2.2){$\mathcal{S}_j$}
\put(92.5, 2.2){$\mathcal{S}_n$}

\put(0.5, 15){0.85}
\put(17.5, 16){0.8}
\put(26, 21){0.9}
\put(43, 10){0.95}
\put(49, 22){0.9}
\put(80, 20){0.91}

\end{overpic}
\vspace{-10pt}
\caption{Matches of an entity and a sub-string}
\label{fig:bipartite-graph}
\end{figure}

To tolerate typos inside tokens, character-based edit-distance is applied
before the Jaccard similarity is applied to measure the similarity between an entity and a sub-string candidate.
The abovementioned Jaccard similarity is called Fuzzy Jaccard
which was first studied by Wang et al.~\cite{wang2011fast} in the context of the string similarity join problem~\cite{wang2010trie}.
Computing the Fuzzy Jaccard similarity is much more complicated,
since one text token may match multiple tokens of an entity and vice versa.
One text token matching to multiple entity tokens frequently occurs
especially when the token edit similarity threshold $\tau$ is small.
Figure~\ref{fig:bipartite-graph} shows a scenario where tokens have multiple matches.
In the figure, a token is represented by a vertex and a match is represented by an edge.
The number next to an edge represents the edit similarity between the two tokens at both ends of the edge.
For example, the edit similarity between $\mathcal{E}_1$ and $\mathcal{S}_1$ is 0.85.
As we can see from the figure, four tokens $\mathcal{E}_1$, $\mathcal{E}_2$, $\mathcal{E}_m$ and $\mathcal{S}_3$
match multiple tokens when the edit similarity threshold $\tau$ is 0.8.
In entity extraction applications, an entity token can match at most one token of a sub-string candidate and vice versa,
so the extra matches should be removed and at most one match is kept for each entity or text token.
We call those extra matches \textbf{redundant matches}.

The \textit{maximum weight matching} algorithm~\cite{west2001introduction} can be applied to remove the redundant matches
before computing the Fuzzy Jaccard similarity.
(Dice similarity and cosine similarity can also use this approach to removing redundant matches.)
Specifically, the maximum weight matching algorithm finds a graph, denoted by $G$, that has the following two properties:
(i) any two edges in $G$ have no common vertex;
(ii) the sum of edit similarity of edges in $G$ is maximum.

After removing the redundant matches, the Fuzzy Jaccard similarity of $\M{E}$ and $\M{S}$ can be computed using the following equation.
\begin{small}
\EQSPACE
\begin{equation*}
FJ = \frac{\sum_{e \in \mathcal{E}', s \in \mathcal{S}'}{eds(e, s)}}
{|\mathcal{E}| + |\mathcal{S}| - \sum_{e \in \mathcal{E}', s \in \mathcal{S}'}{eds(e, s)}}
\end{equation*}
\end{small}%
where $\M{E}' \subseteq \M{E}$ and $\mathcal{S}' \subseteq \M{S}$;
$\M{E}'$ and $\mathcal{S}'$ are subsets of tokens that have matches after removing the redundant matches.
Note that when the edit similarity threshold $\tau$ is one, the above equation is equivalent to Equation~\eqref{eq:jac}.

By considering the weights of tokens (cf. Section~\ref{paper:weighting}),
we can write the Fuzzy Jaccard similarity as follows.
\begin{small}
\begin{equation}
FJ = \frac{\frac{1}{2}\sum_{e \in \mathcal{E}', s \in \mathcal{S}'}{eds(e, s)\cdot (w(e) + w(s))}}
{1 + 1 - \frac{1}{2}\sum_{e \in \mathcal{E}', s \in \mathcal{S}'}{eds(e, s)\cdot (w(e) + w(s))}}
\label{eq:fuzzy-jaccard}
\end{equation}
\end{small}%

\subsubsection{Comparison on the two similarity functions}
\label{paper:fj-analysis}
Computing the Fuzzy Jaccard similarity is expensive.
This is because before computing the similarity, we need to perform the expensive maximum weight matching algorithm
with a time complexity of $\mathcal{O}(m^2n^2)$~\cite{bertsekas1993simple},
where $m$ and $n$ are the number of tokens of the entity $\mathcal{E}$ and that of the sub-string candidate $\mathcal{S}$, respectively.
In comparison, FuzzyED only has a time complexity of $\mathcal{O}(mn)$.

In the following two subsections,
we explain the sub-string candidate producing techniques.
These sub-string candidate producing techniques can be used in the Producing Candidates component
of our algorithm (cf. Figure~\ref{fig:framework}).

\subsection{Producing candidates by enumeration}

For each entity, we can obtain the sub-string candidates by enumerating the results
produced by Li et al.'s algorithm (cf. Figure~\ref{fig:matched-entity-doc})
as we discussed in Section~\ref{paper:matching}.
That is sub-string candidates with one token matching the entity, with two tokens matching the entity,
with three tokens matching the entity, etc.
The number of sub-string candidates produced by this enumeration is of $\mathcal{O}(k^2)$ complexity
and is $\frac{k(1 + k)}{2}$ to be more precise,
where $k$ is the number of text tokens (in the document) that match the entity.
Among the $\frac{k(1 + k)}{2}$ sub-strings,
many of them tend to be unpromising sub-string candidates.
For example, a sub-string candidate with only one matched token is unlikely to match an entity of ten tokens
with the entity similarity threshold $\delta=0.8$.
To generate fewer unpromising sub-string candidates,
we give an approach that only needs to consider sub-string candidates
with the number of matched entity tokens in the range $[l, u]$.
We refer the number of matched tokens of sub-string candidates in the range $[l, u]$ to \textbf{valid matching length}.
The intuition of the valid matching length is that sub-string candidates with too few or too many matched tokens
will not match the entity with the entity edit similarity threshold $\delta$.

In what follows, we first present two propositions for the minimum and maximum valid matching length.
Then, we give details of computing the minimum and maximum valid matching length
for FuzzyED and Fuzzy Jaccard.
Finally, we provide analysis to this enumeration-based candidate producing technique.

\subsubsection{Two propositions of the valid matching length}
For ease of presentation,
we classify the text tokens of a sub-string candidate $\M{S}$ into the following three subsets.
(1) \textit{Unmatched text tokens} denoted by $\hat{\M{S}}$: the text tokens do not match any entity token.
(2) \textit{Redundant matched text tokens} denoted by $\M{S}''$: the text tokens match the entity tokens
but are finally removed (by the maximum weight matching algorithm in Fuzzy Jaccard or by deletion in FuzzyED).
(3) \textit{Valid matched text tokens} denoted by $\M{S}'$:
the text tokens match the entity tokens and are not redundant.
Please note that only the redundant matched text tokens and valid matched text tokens are in the results produced by Li et al.'s algorithm.

\textit{The minimum valid matching length $l$}:
Suppose a sub-string candidate $\mathcal{S}$ has only $l$ tokens that match the entity $\M{E}$,
i.e. the similarity of $\mathcal{S}$ and $\mathcal{E}$ is not smaller than $\delta$.
If $l$ is the minimum valid matching length, the following proposition must be true:
\begin{condition}
All the $l$ text tokens are (i) exactly matched to some entity tokens and (ii) valid matched text tokens.
\label{condition1}
\end{condition}
The proof is straightforward and hence omitted.
According to the proposition, we have $\mathcal{S} = \mathcal{S}'$ and $T_w(\mathcal{S}) = T_w(\mathcal{S}') = 1$
(cf. Equations~\eqref{eq:total_weight}).
Please recall that $\mathcal{S}'$ denotes all the valid matched tokens in $\M{S}$.
Given that $\M{S}$ has the minimum number $l$ of matched tokens,
the above proposition guarantees the maximum possible similarity between $\M{S}$ and $\M{E}$.

\textit{The maximum valid matching length $u$}:
Suppose a sub-string candidate $\mathcal{S}$ has $u$ tokens matched the entity,
i.e. the similarity between $\mathcal{S}$ and $\mathcal{E}$ is not smaller than $\delta$.
If $u$ is the maximum valid matching length, the following proposition must be true:
\begin{condition}
All the tokens of the entity are exactly matched.
\label{condition2}
\end{condition}
The proof is straightforward and hence omitted.
From the above proposition, we have $T_w(\mathcal{E}) = T_w(\mathcal{E}') = 1$ (cf. Equations~\eqref{eq:total_weight}), where $\mathcal{E}'$ denotes all the matched tokens in $\mathcal{E}$.
The above proposition guarantees that (i) the number of valid matched tokens is maximised
(note that the maximum number of valid matched tokens equals to the number of tokens of the entity.) and
(ii) the similarity between $\M{S}$ and $\M{E}$ is maximised given $u$ matched tokens.
\eat{When $\M{S}$ has the maximum number of matched tokens,
some tokens in $\M{S}$ are redundant matched text tokens
while the other tokens in $\M{S}$ are valid matched tokens.
}

\subsubsection{Computing $l$ and $u$ for FuzzyED}
\label{paper:2ED-em}

\textbf{The minimum valid matching length $l$}:
According to Proposition~\ref{condition1}, the substitution and deletion cost are zero,
and only the insertion cost is involved in transforming $\mathcal{S}$ to $\mathcal{E}$.
Therefore, the total cost $FED(\mathcal{E}, \mathcal{S})$ equals to
the insertion cost $T_w(\mathcal{E} \setminus \mathcal{E}')$
where $\mathcal{E} \setminus \mathcal{E}'$ is a subset of the tokens in $\mathcal{E}$
needed to be inserted to $\mathcal{S}$.
According to Equation~\eqref{eq:2ED-sim}, the similarity score is $1 - T_w(\mathcal{E} \setminus \mathcal{E}')$.
When $\M{E}$ and $\M{S}$ are matched, their similarity score is not smaller than $\delta$.
So, we have
\begin{small}
\begin{equation*}
1 - T_w(\mathcal{E} \setminus \mathcal{E}') \geq \delta.
\end{equation*}
\end{small}%
As we know that $T_w(\M{E}) = 1$ (cf. Equation~\eqref{eq:normalise} and~\eqref{eq:total_weight}),
the left part of the above constraint equals to $T_w(\M{E}')$ (i.e. the total weight of the matched entity tokens).
So, we have $T_w(\mathcal{E}') \geq \delta$.

To compute $l$, we add the entity token with the largest weight,
the second largest weight, the third largest weight and so on to $\M{S}'$ until 
the total weight of the tokens in $\M{S}'$ is not smaller than $\delta$.
Then $l$ is computed by $l = |\M{S}'|$.

\textbf{The maximum valid matching length $u$}:
According to Proposition~\ref{condition2}, no insertion cost and no substitution cost are involved;
the only cost is deletion on the redundant matched text tokens.
Since the sub-string candidate should match the entity,
the total weight of the valid matched text tokens,
i.e. $T_w(\mathcal{S}')$, should satisfy the constraint $T_w(\mathcal{S}') \geq \delta$.
From Equations~\eqref{eq:normalise} and~\eqref{eq:total_weight}, we have
\begin{small}
\EQSPACE
\begin{equation}
\displaystyle T_w(\mathcal{S}') = \frac{T_{idf}(\mathcal{S}')}{T_{idf}(\mathcal{S})}.
\label{eq:total_matched_weight}
\EQSPACE
\end{equation}
\end{small}%
Recall that $\M{S}$ is a sub-string candidate and $\M{S}'$ is the valid matched tokens in $\M{S}$ (i.e. $\M{S}' \subseteq \M{S}$).
Except the valid matched tokens in $\M{S}'$, the sub-string candidate $\M{S}$ also contains redundant matched text tokens $\mathcal{S}''$,
unmatched text tokens $\hat{\mathcal{S}}$.
We can rewrite Equation~\eqref{eq:total_matched_weight} in the following form.
\begin{small}
\EQSPACE
\begin{equation*}
\displaystyle T_w(\mathcal{S}') = \frac{T_{idf}(\mathcal{S}')}{T_{idf}(\mathcal{S}') + T_{idf}(\mathcal{S}'') + T_{idf}(\hat{\mathcal{S}})}
\end{equation*}
\end{small}%
Since we compute the maximum valid matching length $u$ of the matched tokens in a sub-string candidate,
we only know all the \textit{matched} text tokens (cf. Section~\ref{paper:matching}) to the entity.
So we write the above equation in the following form.
\begin{small}
\EQSPACE
\begin{equation*}
\frac{T_{idf}(\mathcal{S}')}{T_{idf}(\mathcal{S}') + T_{idf}(\mathcal{S}'')} \geq \displaystyle T_w(\mathcal{S}')
\end{equation*}
\end{small}%
As {\small $\displaystyle T_w(\mathcal{S}') \geq \delta$},
we have 
\begin{small}
\EQSPACE
\begin{equation}
\frac{T_{idf}(\mathcal{S}')}{T_{idf}(\mathcal{S}') + T_{idf}(\mathcal{S}'')} \geq \delta.
\label{eq:finding-u}
\end{equation}
\end{small}%

We know that $u$ equals to the number of the tokens of $\mathcal{S}'$ plus the number of tokens in $\M{S}''$.
From Proposition~\ref{condition2}, $T_{idf}(\mathcal{S}')$ equals to
$T_{idf}(\mathcal{E})$ and is a constant.
The number of the tokens in $\mathcal{S}''$ is maximised when each redundant token has the smallest IDF value.
Therefore, $u$ is the maximum number when all the tokens of $\M{S}''$ match $\M{E}$'s token with the smallest IDF value.
To compute $u$, we keep adding the same entity token (the one with the smallest IDF value among the tokens of $\M{E}$)
to $\M{S}''$ until Inequality~\eqref{eq:finding-u} does not hold.
Then $u$ is computed by $u = |\M{S}'| + |\M{S}''|$.

\subsubsection{Computing $l$ and $u$ for Fuzzy Jaccard}
For a sub-string candidate to match an entity,
the similarity of the sub-string candidate and the entity must satisfy the condition $\displaystyle FJ \geq \delta$
where $FJ$ is computed using Equation~\eqref{eq:fuzzy-jaccard}.
Combining Equation~\eqref{eq:fuzzy-jaccard} and $\displaystyle FJ \geq \delta$,
we have
\begin{small}
\EQSPACE
\begin{equation*}
\sum_{e \in \mathcal{E}', s \in \mathcal{S}'}{eds(e, s)\cdot (w(e) + w(s))} \geq \frac{4\delta}{1 + \delta}.
\end{equation*}
\end{small}%
As $eds(e, s) \leq 1$, we let $eds(e, s) = 1$. 
Then we have
\begin{small}
\EQSPACE
\begin{equation*}
\displaystyle \sum_{e \in \M{E}', s \in \M{S}'}{(w(e) + w(s))} 
= \sum_{e \in \M{E}'}{w(e)} + \sum_{s \in \M{S}'}{w(s)}
\geq \frac{4\delta}{1 + \delta}.
\EQSPACE
\end{equation*}
\end{small}%
Using Equation~\eqref{eq:total_weight},
we can rewrite the above inequality as 
\begin{small}
\EQSPACE
\begin{equation}
\displaystyle T_w(\mathcal{E}') + T_w(\mathcal{S}') \geq \frac{4\delta}{1 + \delta}.
\label{eq:fj-constraint-2}
\end{equation}
\end{small}%

\textbf{The minimum valid matching length $l$}:
According to Proposition~\ref{condition1}, we have $T_w(\mathcal{S}') = 1$.
We can rewrite Inequality~\eqref{eq:fj-constraint-2} by putting $T_w(\mathcal{S}') = 1$ into it and
we have
\begin{small}
\EQSPACE
\begin{equation*}
\displaystyle T_w(\mathcal{E}') \geq \frac{3\delta - 1}{1 + \delta}.
\end{equation*}
\end{small}%
Computing $l$ here is identical to the process of computing $l$ in FuzzyED,
except the threshold here is $\frac{3\delta - 1}{1 + \delta}$ instead of $\delta$.

\textbf{The maximum valid matching length $u$}:
According to Proposition~\ref{condition2}, we have $T_w(\mathcal{E}') = 1$.
By putting $T_w(\mathcal{E}') = 1$ into Inequality~\eqref{eq:fj-constraint-2}, we have 
\begin{small}
\EQSPACE
\begin{equation}
T_w(\mathcal{S}') \geq \frac{3\delta - 1}{1 + \delta}.
\label{eq:fj-constraint-max}
\EQSPACE
\end{equation}
\end{small}%

Following the same process of deriving from Equation~\eqref{eq:total_matched_weight} to Inequality~\eqref{eq:finding-u},
we can rewrite Inequality~\eqref{eq:fj-constraint-max} as follows.
\begin{small}
\EQSPACE
\begin{equation*}
\displaystyle \frac{T_{idf}(\mathcal{S}')}{T_{idf}(\mathcal{S}') + T_{idf}(\mathcal{S}'')} \geq \frac{3\delta - 1}{1 + \delta}.
\EQSPACE
\end{equation*}
\end{small}%

Then, computing $u$ here is identical to the process of computing $u$ in FuzzyED,
except the threshold here is $\frac{3\delta - 1}{1 + \delta}$ instead of $\delta$.

\subsubsection{Analysis of producing candidates by enumeration}
\label{paper:em-analysis}

In the enumeration-based candidate producing technique,
the number of sub-string candidates generated using the valid matching length is of $\mathcal{O}(k)$ complexity
and is $(u - l) \times k$ to be more precise,
where $k$ is the number of matched tokens in the document.
Even though we have reduced the number of sub-string candidates from $\frac{k(1 + k)}{2}$
to $(u - l) \times k$,
many unpromising sub-string candidates are generated and require measuring the two-level similarity (e.g. FuzzyED).
Next, we propose a novel spanning-based candidate producing technique that
reduces the number of sub-string candidates which requires measuring the two-level similarity to $k$.

\subsection{Producing candidates by spanning}
\label{paper:2ED-s}
We notice that the large number of unpromising sub-string candidates generated by the enumeration-based
candidate producing technique is because many matched tokens are not important tokens
(i.e. tokens with small IDF values such as stop words~\cite{dhillon2001co}).
Those tokens are likely to appear many times in a document and result in generating many unpromising sub-string candidates.
Here, we propose a spanning-based candidate producing technique
that makes use of important tokens which we call \textbf{core tokens}.
The technique starts from a core token and uses left and right spanning to find sub-string candidates for measuring the two-level similarity.
To determine the left and right boundaries of a sub-string candidate,
we design a lower bound dissimilarity derived from the two-level similarity.

In what follows, we first present the technique to find core tokens.
Then, we provide the key steps of our spanning-based candidate producing technique.
After that, we give details of the spanning-based candidate producing technique to FuzzyED and Fuzzy Jaccard.
Lastly, we design techniques for reusing computation in spanning-based candidate producing,
and analyse the candidate producing technique.

\subsubsection{Finding core tokens of an entity}
\label{paper:2ED-core}
As we have discussed in Section~\ref{paper:weighting}, each token is associated with a weight.
The weights of tokens can help reduce the number of unpromising sub-string candidates.
Our key idea is to find a subset of entity tokens (i.e. core tokens) to represent the entity.
For instance, we may use core tokens \{University, Oxford\}
to represent the entity with tokens \{The, University, of, Oxford\}.
The rest of the tokens with smaller weights, such as \{The, of\} in the example, are called \textbf{optional tokens} in this paper.

Formally, given an entity similarity threshold $\delta$ and an entity with $m$ tokens
$\mathcal{E}=\{\mathcal{E}_1, \mathcal{E}_2, ..., \mathcal{E}_m\}$,
we construct a set $\mathcal{C}$ of $q$ tokens to represent the entity $\mathcal{E}$ where $\mathcal{C} \subseteq \mathcal{E}$.
The remaining $(m - q)$ tokens in $\mathcal{E}\setminus \mathcal{C}$ form a set $O$ corresponding to the optional tokens.
The property of core tokens is that at least one core token should appear in a sub-string candidate to allow the candidate to match the entity.
Next, we first describe the approaches to finding core tokens in the settings
of using the FuzzyED and Fuzzy Jaccard similarity.
Then, we provide more details of the properties of core tokens.

\subsubsection*{Core tokens for FuzzyED}

The core token set $\mathcal{C}$ should satisfy the following constraint.
\begin{small}
\EQSPACE
\begin{equation}
T_w(\mathcal{C}) > 1 - \delta
\label{eq:core-ed}
\end{equation}
\end{small}%
The above constraint guarantees that the total weight of tokens
in the optional token set $O$ (where $O = \mathcal{E}\setminus \M{C}$) to be smaller than $\delta$,
because {\small $T_w(O) = 1 - T_w(\M{C}) < \delta$}.

\subsubsection*{Core tokens for Fuzzy Jaccard}

Similar to FuzzyED, the core token set $\mathcal{C}$ in Fuzzy Jaccard should satisfy the following constraint.
\begin{small}
\EQSPACE
\begin{equation}
T_w(\mathcal{C}) >\frac{2(1-\delta)}{1 + \delta}
\label{eq:core-fj}
\EQSPACE
\end{equation}
\end{small}%
Due to the space limitation, we omit the details of deriving the above inequality.
The above constraint guarantees that the similarity of $\M{S}$ and $\M{E}$
is smaller than $\delta$,
given that not any core token is matched.

\subsubsection*{Properties of core tokens}

The following lemma shows that at least one core token should appear in a sub-string candidate
to allow the sub-string candidate to match the entity.
\begin{lemma}
Given a sub-string candidate $\mathcal{S}$ that matches an entity $\mathcal{E}$ (i.e. the similarity between $\mathcal{S}$ and $\mathcal{E}$ is not smaller than $\delta$),
the sub-string candidate $\mathcal{S}$ must have at least one text token matching to a core token of the entity $\mathcal{E}$.
\label{lemma-core}
\end{lemma}

The proof to the lemma can be found in Appendix~\ref{appendix:core-proof}.
According to the above lemma, the sub-string candidates not containing any core token can be discarded without sacrificing recall. 
Hence, core tokens are good starting points to find the sub-string candidates.

Note that the number of core tokens of an entity should be as small as possible,
because a core token may match many text tokens;
those matched text tokens may generate many unpromising sub-string candidates which require measuring the two-level similarity.
To minimise the number of core tokens to represent an entity (i.e. minimising the cardinality $q$ of $\mathcal{C}$),
we select $q$ tokens with the largest weights from $\mathcal{E}$
to make $\mathcal{C}$ just satisfy the constraint for core tokens, e.g. Constraint~\eqref{eq:core-ed} for FuzzyED.
\eat{An important property of the core token and the optional tokens is $w(e_i) \geq w(e_j)$
where $\forall e_i \in \mathcal{C}$ and $\forall e_j \in O$.}

In what follows, we explain the key steps of producing a candidate starting from a core token
by left and right spanning.

\subsubsection{The spanning process of producing a candidate}
Since the core tokens represent the entity,
we only use the core tokens as query tokens to find their matching positions in the document
using Li et al.'s algorithm.
The matched results of the entity in the document are similar to the results shown in Figure~\ref{fig:matched-entity-doc}.
In many cases, the left and right boundaries of a sub-string candidate are not core tokens.
Hence, we need to check the left (right) side of the leftmost (rightmost) core token in the sub-string candidate
and see if any optional tokens near the core token can be included into the sub-string candidate.
We call the process of finding the left (right) boundary of the sub-string candidate \textit{left spanning} (\textit{right spanning}).
To determine when the spanning should be terminated,
we compute a lower bound of the dissimilarity between the sub-string candidate and the entity.
When the left spanning or right spanning results in the lower bound dissimilarity higher than the threshold $(1 - \delta)$,
the spanning should be terminated.

\begin{figure}
\hspace{15pt}
\begin{overpic}[width=2.2in, height=1.6in]
{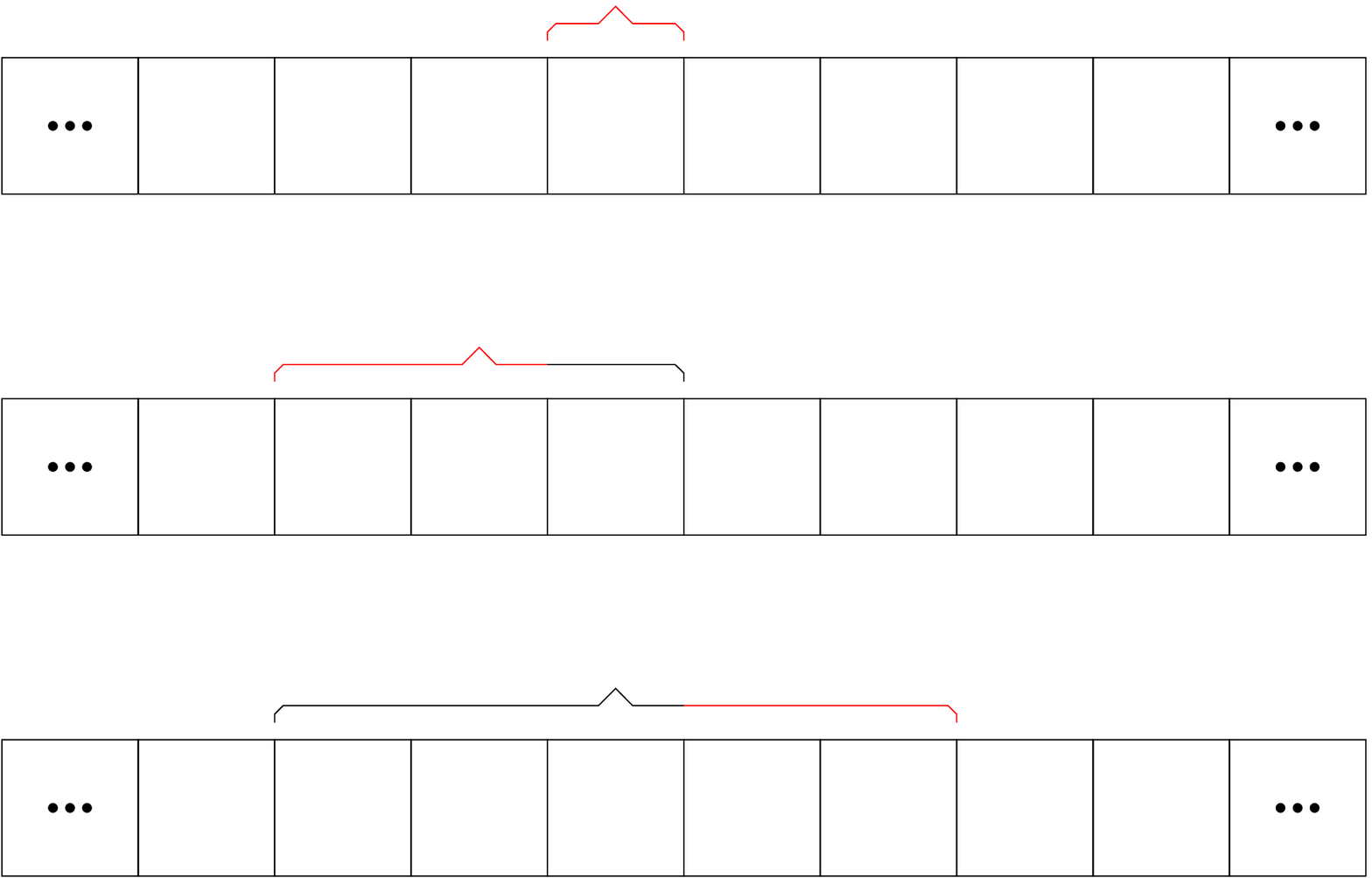}

\put(26, 73){current sub-string}
\put(103, 60){initial}
\put(13, 60){X}
\put(23, 60){X}
\put(33, 60){X}
\put(44, 60){$\mathcal{C}_1$}
\put(53, 60){$\mathcal{C}_2$}
\put(63, 60){X}
\put(73.5, 60){X}
\put(84, 60){$\mathcal{C}_1$}

\put(16, 45){current sub-string}
\put(108, 34.5){left}
\put(103, 30.5){spanning}
\put(13, 32){X}
\put(23, 32){$O_1$}
\put(33, 32){X}
\put(44, 32){$\mathcal{C}_1$}
\put(53, 32){$\mathcal{C}_2$}
\put(63, 32){X}
\put(73.5, 32){X}
\put(84, 32){$\mathcal{C}_1$}

\put(26, 17){current sub-string}
\put(107, 7){right}
\put(103, 3){spanning}
\put(13, 4){X}
\put(23, 4){$O_1$}
\put(33, 4){X}
\put(44, 4){$\mathcal{C}_1$}
\put(53, 4){$\mathcal{C}_2$}
\put(63, 4){$O_3$}
\put(73.5, 4){X}
\put(83, 4){$\mathcal{C}_1$}

\end{overpic}
\vspace{-5pt}
\caption{Spanning from the core token}
\label{fig:spanning}
\end{figure}

Figure~\ref{fig:spanning} shows an overview of the process of finding the boundaries of a sub-string candidate.
Initially, the sub-string which we call \textit{current sub-string} has only one token (i.e. the core token $\mathcal{C}_1$).
Then, the left spanning leads to an optional token $O_1$ included in the current sub-string.
The left spanning is terminated because of the lower bound dissimilarity is higher than $(1-\delta)$
if more tokens in the left side are included.
By right spanning, the current sub-string covers one more core token (i.e. $\mathcal{C}_2$)
and one optional token (i.e. $O_3$).
The current sub-string cannot be further extended because of the high lower bound dissimilarity,
and hence we obtain the sub-string candidate which requires measuring the two-level similarity.

In what follows, we first present the intuition of computing the lower bound dissimilarity.
Then, we describe the key ideas of the left spanning and the right spanning.
We postpone the presentation on more details of producing the sub-string candidates
specifically for FuzzyED and Fuzzy Jaccard until Section~\ref{paper:spanning-fj} and Section~\ref{paper:spanning-ed}.

\textbf{The lower bound dissimilarity}:
As demonstrated in Figure~\ref{fig:spanning},
we start from a sub-string with a core token, and then extend the sub-string by left and right spanning.
Spanning the current sub-string to include a nearby token raises change to the similarity score.
To determine when the left/right spanning process should be terminated,
we compute the lower bound dissimilarity for the current sub-string with the nearby token included.
We denote the lower bound similarity by $\mathcal{B}_\perp$,
the computing of which depends on the specific similarity function (e.g. FuzzyED).
\eat{
If the lower bound dissimilarity is smaller than $(1-\delta)$,
the nearby token is included into the current sub-string;
otherwise, the left/right spanning process is terminated.
}

\eat{Without loss of generality, we suppose the text token $t$ matches the $i^{th}$ token of the entity $\mathcal{E}$,
so we set the $i^{th}$ element of the array $M$ to the edit similarity of $t$ and $\mathcal{E}_i$
(i.e. $M_i = eds(t, \mathcal{E}_i)$).
}

\textbf{Left spanning}:
Here, we provide the details of extending the current sub-string via left spanning.
Another interpretation to the left spanning is to find the left boundary of the sub-string candidate.
To begin with, we start from the first matched text token
(e.g. the first $\mathcal{C}_1$ in Figure~\ref{fig:spanning}) in the document.
Then, we span to the left side of the current sub-string by one text token, denoted by $t$.
Next, we compute the lower bound dissimilarity $\LB$.
If $\LB$ is smaller than $(1-\delta)$, we span the current sub-string to cover the text token $t$;
otherwise, the left spanning is terminated.
When the left spanning is terminated, the leftmost matched text token is identified as the left boundary of the sub-string candidate.

\textbf{Right spanning}:
After the left spanning, we span the current sub-string to include the tokens to its right side.
The right spanning is identical to the left spanning
and hence is not discussed afterwards.

Next, we describe the details of computing the lower bound dissimilarity and the left spanning process
for FuzzyED and Fuzzy Jaccard.

\subsubsection{Producing a candidate for FuzzyED}
\label{paper:spanning-ed}

\textbf{The lower bound dissimilarity}:
In the setting of the FuzzyED similarity,
the lower bound dissimilarity $\LB$ is from the total deletion cost and the total substitution cost
while producing the sub-string candidate.
Please note that the lower bound insertion cost is always zero,
because all the entity tokens potentially have exact matches by left and right spanning.
To compute the lower bound dissimilarity $\mathcal{B}_\perp$ more efficiently,
we maintain the total IDF values $\mathcal{V}_T$ for all the tokens in the current sub-string,
and the total IDF values $\mathcal{V}_R$ for those text tokens needed to be deleted from the current sub-string.
$\mathcal{V}_T$ is initialised to the IDF value of the core token and $\mathcal{V}_R$ is initialised to 0.

The substitution cost between two similar tokens {\small $\mathcal{E}_i$} and {\small $\mathcal{S}_j$} is
{\small $(1 - eds(\mathcal{E}_i, \mathcal{S}_j))\times (w(\mathcal{E}_i) + w(\mathcal{S}_j))$}
according to Equation~\eqref{eq:substitution}.
We cannot simply include the substitution cost into the lower bound dissimilarity,
as there may exist another not included token {\small $\mathcal{S}'_r$} that is
more similar to {\small $\mathcal{E}_i$} than {\small $\mathcal{S}_j$},
i.e. {\small $eds(\mathcal{E}_i, \mathcal{S}'_r) > eds(\mathcal{E}_i, \mathcal{S}_j)$}.
If such {\small $\mathcal{S}'_r$} exists, we need to delete {\small $\mathcal{S}_j$}
with cost {\small $w(\mathcal{S}_j)$} later in the spanning.
Note that the substitution cost {\small $(1 - eds(\mathcal{E}_i, \mathcal{S}_j))\times (w(\mathcal{E}_i) + w(\mathcal{S}_j))$}
may be larger than the deletion cost {\small $w(\mathcal{S}_j)$}.
Hence, the lowest cost of including {\small $\mathcal{S}_j$} to the current sub-string
is set to {\small $(1 - eds(\mathcal{E}_i, \mathcal{S}_j)) \times w(\mathcal{S}_j)$}
which is smaller than both {\small $(1 - eds(\mathcal{E}_i, \mathcal{S}_j))\times (w(\mathcal{E}_i) + w(\mathcal{S}_j))$}
and {\small $w(\mathcal{S}_j)$}.
The lowest cost of including the text token can be represented in the form of IDF values by
{\small $(1 - eds(\mathcal{E}_i, \mathcal{S}_j))\times idf(\mathcal{S}_j)$}.
This lowest cost of including the text token is equivalent to deleting a token
with an IDF value of {\small $(1 - eds(\mathcal{E}_i, \mathcal{S}_j))\times idf(\mathcal{S}_j)$}.
In what follows, we compute the lower bound dissimilarity as if we only considered deletion cost.

For ease of computing the lower bound dissimilarity,
we maintain an array {\small $M$} with the length of {\small $|\mathcal{E}|$}.
The $i^{th}$ element of the array, denoted by {\small $M_i$} and {\small $i \in \{i: \tau \leq M_i \leq 1\}$},
corresponds to the edit similarity between the most similar text token of the current sub-string 
and the $i^{th}$ entity token of {\small $\mathcal{E}$} (i.e. {\small $\mathcal{E}_i$}).
Note that some elements (e.g. {\small $M_j$}) in {\small $M$} are marked as none
if the corresponding entity tokens have no matched text token
(e.g. no token in the current sub-string {\small $\M{S}$} matches {\small $\M{E}_j$}).
The equation of computing the lower bound dissimilarity can be expressed as follows.
\begin{small}
\EQSPACE
\begin{equation}
\mathcal{B}_\perp = \frac{\mathcal{V}_R + \sum_i{(1 - M_i) \times idf({\mathcal{S}'_i})}}
{\mathcal{V}_T + \sum_r{idf(\mathcal{E}_r)}}
\label{eq:lower-bound}
\EQSPACE
\end{equation}
\end{small}%
where {\small $i \in \{i: \tau \leq M_i \leq 1\}$} and {\small $r \in \{r: M_r < 1\}$}.

The numerator of Equation~\eqref{eq:lower-bound} represents the total ``deletion cost'':
true deletion cost $\mathcal{V}_R$ and the substitution cost $\sum_i{(1 - M_i) \times idf({\mathcal{S}'_i})}$
where $\mathcal{S}'_i$ is the text token that is the most similar to $\mathcal{E}_i$.
The denominator is the ideal total IDF value of the sub-string;
$\mathcal{V}_T$ is the total IDF value of the current sub-string;
the term $\sum_r{idf(\mathcal{E}_r)}$ of the denominator is the total IDF value of all the not exactly matched entity tokens.
We can prove that the lower bound dissimilarity increases monotonically as the sub-string spans.
The key idea of the proof is that adding the same value $idf(t) > 0$
to the numerator and the denominator of Equation~\eqref{eq:lower-bound}
leads to the value of $\mathcal{B}_\perp$ increasing.

\textbf{Left spanning}:
For updating $\mathcal{V}_T$ and $\mathcal{V}_R$ in this spanning, we need to handle the following two cases separately.
Suppose the token to the left side of the current sub-string is $t$,
and $t$ is included into the sub-string after spanning.
\begin{itemize}
\item Case 1: $t$ does not match any tokens of {\small $\mathcal{E}$}, so we need to delete $t$.
			  Hence, we update {\small $\mathcal{V}_R$} by {\small $\mathcal{V}_R = \mathcal{V}_R + idf(t)$}, and
			  we update the total IDF value {\small $\mathcal{V}_T$} by {\small $\mathcal{V}_T = \mathcal{V}_T + idf(t)$}.

\item Case 2: $t$ matches a token {\small $\mathcal{E}_j$} of {\small $\mathcal{E}$}.
			  We update {\small $\mathcal{V}_T$} by {\small $\mathcal{V}_T = \mathcal{V}_T + idf(\mathcal{E}_j)$}.
			  We consider this as a substitution operation and update {\small $\mathcal{V}_R$} by
			  the following two cases.
			\begin{itemize}
				\item No other text token in the current sub-string matches {\small $\mathcal{E}_j$}.
					  We update {\small $M_j$} by {\small $M_j = eds(t, \mathcal{E}_j)$},
					  and we do not update {\small $\mathcal{V}_R$} due to no deletion required.
				\item A text token in the current sub-string has matched to {\small $\mathcal{E}_j$}.
					  We update {\small $\mathcal{V}_R$} by {\small $\mathcal{V}_R = \mathcal{V}_R + idf(t)$},
					  and {\small $M_j$} by {\small $M_j = \max\{M_j, eds(t, \mathcal{E}_j)\}$}.
			\end{itemize}
\end{itemize}
After the update of $\mathcal{V}_T$, $\mathcal{V}_R$ and $M$,
we compute the lower bound dissimilarity $\LB$ using Equation~\eqref{eq:lower-bound}.
When $\LB > 1 - \delta$, the left spanning terminates.

\subsubsection{Producing a candidate for Fuzzy Jaccard}
\label{paper:spanning-fj}

\textbf{The lower bound dissimilarity}:
For Fuzzy Jaccard, in order to compute the lower bound dissimilarity $\LB$ of the current sub-string $\M{S}$,
we first compute the maximum possible similarity score $\SC$ of the sub-string,
and then $\LB = (1 - \SC)$.
According to Equation~\eqref{eq:fuzzy-jaccard},
the similarity score of $\M{E}$ and $\M{S}$ reaches the maximum value when the term
$\sum_{e \in \mathcal{E}', s \in \mathcal{S}'}{eds(e, s)\cdot (w(e) + w(s))}$ is maximised.
We can rewrite the term in the following form.
\begin{small}
\EQSPACE
\begin{equation*}
\sum_{e \in \mathcal{E}', s \in \mathcal{S}'}{eds(e, s)\cdot w(e)} +
\sum_{e \in \mathcal{E}', s \in \mathcal{S}'}{eds(e, s)\cdot w(s)}
\EQSPACE
\end{equation*}
\end{small}%
The current sub-string $\M{S}$ has the maximum similarity to the entity $\M{E}$,
when all the entity tokens are exactly matched.
That is $\sum_{e \in \M{E}', s \in \M{S}'}{eds(e, s)\cdot w(e)} = 1$. So we have
\begin{small}
\EQSPACE
\begin{equation*}
1 + \sum_{e \in \mathcal{E}', s \in \mathcal{S}'}{eds(e, s)\cdot w(s)}
\EQSPACE
\end{equation*}
\end{small}%
The above term is maximised when
$\sum_{e \in \mathcal{E}', s \in \mathcal{S}'}{eds(e, s)\cdot w(s)}$ reaches its maximum possible value.
Next, we replace the weight by the IDF values, and we have
\begin{small}
\EQSPACE
\begin{equation}
\sum_{e \in \mathcal{E}', s \in \mathcal{S}'}{eds(e, s)\cdot w(s)}=
\frac{\sum_{e \in \mathcal{E}', s \in \mathcal{S}'}{eds(e, s)\cdot idf(s)}}
{T_{idf}(\M{S})} 
\label{term:idf}
\EQSPACE
\end{equation}
\end{small}%
where $T_{idf}(S)$ is the total IDF value of the current sub-string (cf. Equation~\eqref{eq:tidf}).
The above term is maximised,
when $eds(e, s)$ equals to the edit similarity of the entity token $e$ to the most similar token of the current sub-string.
Recall that the $i^{th}$ element of $M$ is the similarity of $\M{E}_i$ and the most similar token of the current sub-string.
Hence, we can rewrite the term~\eqref{term:idf} in the following form using $M$.
\begin{small}
\EQSPACE
\begin{equation}
\frac{\sum{M_i\cdot idf(\M{S}'_i)}}
{T_{idf}(\M{S})}
\label{eq-jaccard-term}
\EQSPACE
\end{equation}
\end{small}%
where $\M{S}'_i$ is the text token which is the most similar to $\M{E}_i$ in the current sub-string $\M{S}$.
The value of the above term may increase as more tokens are included via the left/right spanning.
The text tokens that improve the similarity score are those similar to the entity tokens
(i.e. through improving the value of $M_i$).
We can modify~\eqref{eq-jaccard-term} to a term that has the maximum value as follows
\begin{small}
\EQSPACE
\begin{equation}
\frac{\sum{M_i\cdot idf(\M{S}'_i)} + \sum_r{(1-M_r)idf(\M{S}'_r)}}
{T_{idf}(\M{S}) + \sum_r{idf(\M{E}_r)}}
\label{term:max-fj}
\EQSPACE
\end{equation}
\end{small}%
where $r \in \{r: M_r < 1\}$,
and the token $\M{S}'_r$ (which is similar to $\M{E}_r$) is added to the above term only when the value of the term ~\eqref{term:max-fj} increases.
Note that the left/right spanning process can be terminated when the term~\eqref{term:max-fj} cannot be increased.
The following lemma identifies the tokens that can increase the value of the term~\eqref{term:max-fj},
and hence increase the Fuzzy Jaccard similarity.

\begin{lemma}
A token $\M{S}'_r$ can increase the Fuzzy Jaccard similarity of the current sub-string $\M{S}$ if
\begin{small}
\EQSPACE
\begin{equation*}
(1-M_r) \geq \frac{\sum{M_i\cdot idf(\M{S}'_i)}}{T_{idf}(\M{S})}
\EQSPACE
\end{equation*}
\end{small}%
\label{lemma-increase}
\vspace{-15pt}
\end{lemma}
The proof of Lemma~\ref{lemma-increase} can be found in Appendix~\ref{appendix:fj-bound}.
Based on Equation~\eqref{eq:fuzzy-jaccard},
the maximum similarity score is computed using the following similarity function.
\begin{small}
\EQSPACE
\begin{equation}
\SC = \frac{\frac{1}{2}(1 + T)}
{1 + 1 - \frac{1}{2}(1 + T)}
\label{eq:jaccard-max}
\EQSPACE
\end{equation}
\end{small}%
where $T$ is the term~\eqref{term:max-fj}.
Then the lower bound dissimilarity can be computed by $\LB = 1 - \SC$.
We can prove that the lower bound dissimilarity of Fuzzy Jaccard increases monotonically.
The proof is straightforward and hence omitted.

\textbf{Left Spanning}:
We can use the lower bound dissimilarity discussed above to determine when the left spanning can be terminated.
We denote {\small $\M{V}_T = T_{idf}(\M{S})$}.
When spanning, we need to update the value of {\small $\M{V}_T$}.
Suppose the token to the left side of the current sub-string is $t$.
We update {\small $\M{V}_T$} by {\small $\M{V}_T = \M{V}_T + idf(t)$}.
The numerator of the term~\eqref{term:max-fj} is handled by the following two cases.
\begin{itemize}
	\item Case 1: $t$ does not match to any entity token. Then the numerator does not need to be updated.
	\item Case 2: $t$ matches to an entity token {\small $\M{E}_i$}.
		\begin{itemize}
			\item {\small $\M{E}_i$} has no matching to any other text token.
			Then, {\small $M_i=eds(\M{E}_i, t)$}. The updated {\small $M_i$} contributes
			to increasing the numerator of term~\eqref{term:max-fj}.
			\item {\small $\M{E}_i$} has other matching to some text tokens in the current sub-string;
			then, {\small $M_i = \max\{M_i, eds(t, \M{E}_i)\}$}.
		\end{itemize}
\end{itemize}
After the update of $\M{V}_T$ and $M$,
we can recompute the maximum similarity using Equation~\eqref{eq:jaccard-max},
and compute the lower bound $\LB$. 
If $\LB > 1 - \delta$, the left spanning should be terminated.

\begin{figure}
\hspace{20pt}
\begin{overpic}[width=2.2in, height=0.8in]
{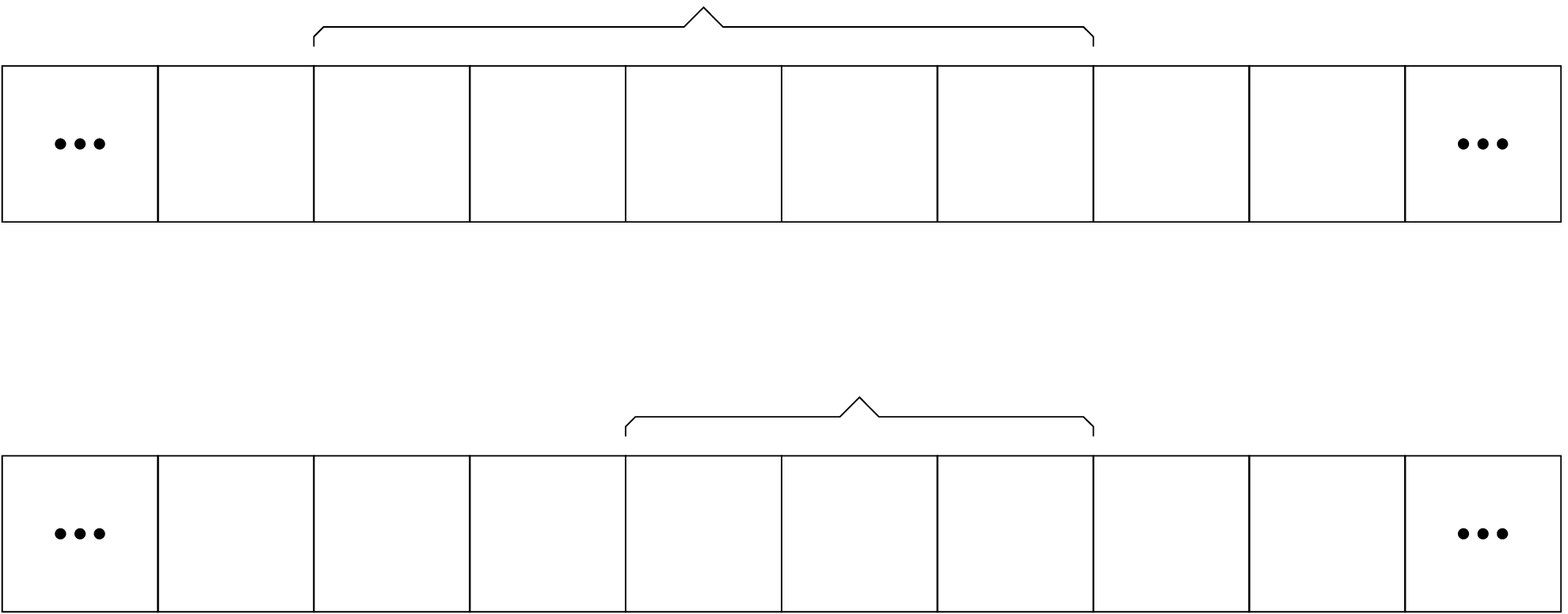}

\put(10, 38){previous sub-string candidate}
\put(106, 29){before}
\put(103, 24){shrinking}
\put(13, 26){X}
\put(22, 26){$O_1$}
\put(33, 26){X}
\put(43, 26){$\mathcal{C}_1$}
\put(52, 26){$\mathcal{C}_2$}
\put(61.5, 26){$O_3$}
\put(72.5, 26){X}
\put(83, 26){$\mathcal{C}_1$}

\put(33, 15){current sub-string}
\put(107, 6){after}
\put(103, 1){shrinking}
\put(13, 3){X}
\put(22, 3){$O_1$}
\put(33, 3){X}
\put(44, 3){$\mathcal{C}_1$}
\put(53, 3){$\mathcal{C}_2$}
\put(61.5, 3){$O_3$}
\put(73.5, 3){X}
\put(83, 3){$\M{C}_1$}

\end{overpic}
\vspace{-8pt}
\caption{Shrinking the previous sub-string}
\label{fig:shrinking}
\vspace{-10pt}
\end{figure}

\subsubsection{Reusing computation in producing candidates}
\label{paper:reuse-comp-shrink}

The boundaries of a sub-string candidate should start and end with matched tokens,
because the leading/ending unmatched tokens of the sub-string candidate are not part of the entity.
We can use this property to reuse some computation while finding the boundaries
of a neighbour sub-string candidate (i.e. the sub-string candidate next to the previously found sub-string candidate in the document).
We refer to text tokens that match the entity $\M{E}$ as \textit{landmark} tokens.

\textbf{Shrinking}: To find the neighbour sub-string candidate,
we shrink the previous sub-string candidate by one landmark token.
That is, the left boundary is moved from the leftmost landmark, denoted by $l_1$,
to the second leftmost landmark, denoted by $l_2$.
Figure~\ref{fig:shrinking} gives an example of shrinking the previous sub-string candidate.
The leftmost landmark $l_1$ and the second leftmost landmark $l_2$ of the previous sub-string candidate
are $O_1$ and $\mathcal{C}_1$, respectively;
after shrinking, we obtain the current sub-string with $\mathcal{C}_1$ as the leftmost landmark.

Suppose $l_1$ matches the $i^{th}$ entity token $\mathcal{E}_i$.
The total IDF value $\mathcal{V}_T$ of the sub-string after shrinking can be updated as follows.
\begin{small}
\begin{equation}
\mathcal{V}_T = \mathcal{V}_T - \mathcal{V}_s - idf(l_1)
\end{equation}
\end{small}%
where $\mathcal{V}_s = \sum{idf(t_j)}$, and $t_j$ is the text token
between the leftmost landmark $l_1$ and the second leftmost landmark $l_2$.
Next, we provide the formulas for updating other values specially for FuzzyED and Fuzzy Jaccard.

\textit{Shrinking for FuzzyED}:
We update $\mathcal{V}_R$ using the following equation.
\begin{small}
\begin{equation*}
\mathcal{V}_R = \begin{cases}
		\mathcal{V}_R - \mathcal{V}_s - idf(l_1)  \qquad  \text{if } eds(l_1, \mathcal{E}_i) < M_i, \\
		\mathcal{V}_R - \mathcal{V}_s - idf(t)  \qquad \qquad \qquad \text{otherwise.}
	  \end{cases}
\end{equation*}
\end{small}%
The first case is for removing the landmark token $l_1$ which is not the most similar token to $\M{E}_i$;
the second case is for removing $l_1$ which is the most similar token to $\M{E}_i$.
If $l_1$ is the most similar token to $\M{E}_i$,
we need to update $M_i$ by $M_i = eds(t, \mathcal{E}_i)$
where $t$ is the second most similar token to $\mathcal{E}_i$ in the previous sub-string candidate.

\textit{Shrinking for Fuzzy Jaccard}:
We let $\M{V}_m = \sum{M_i \cdot idf(\M{S}'_i)}$.
Then $\M{V}_m$ is updated using the following equation.
\begin{small}
\begin{equation*}
\M{V}_m = \begin{cases}
		\M{V}_m   \qquad \qquad \qquad \qquad \quad \text{if } eds(l_1, \M{E}_i) < M_i, \\
		\M{V}_m - (M_i - eds(t, \M{E}_i)) \cdot idf(l_1)  \ \ \text{otherwise.}
	  \end{cases}
\end{equation*}
\end{small}%
In the first case, we do not need to update $\M{V}_m$, since $l_1$ is not part of $\M{V}_m$.
In the later case, we need to find the second most similar token $t$ to $\M{E}_i$,
update $\M{V}_m$ accordingly and update $M_i$ by $M_i = eds(t, \M{E}_i)$.

After the shrinking, we can start the right spanning to find the right boundary of the new sub-string candidate.

\subsubsection{Analysis of producing candidates by spanning}
\label{paper:2ED-s-analysis}
Using the spanning-based technique, the number of sub-strings required measuring the two-level similarity is $k$ at most,
where $k$ is the number of matched tokens (including core tokens and optional tokens).
To understand this, we refer to Figure~\ref{fig:shrinking}.
Every time, we shrink the previous sub-string candidate by one matched text token and find a new sub-string candidate.
Hence, we perform $k$ shrinking at most,
and each shrinking corresponds to a sub-string candidate.
Therefore, the spanning-based candidate producing technique generates $k$ sub-string candidates at most.
In comparison, the enumeration-based candidate producing technique generates $(u - l) \times k$ sub-strings
as we have analysed in Section~\ref{paper:em-analysis}.

\textbf{Not using core tokens}: The spanning-based candidate producing technique can be applied to the case of not using core tokens.
The number of sub-string candidates requires measuring the two-level similarity is also $k$ (i.e. all the matched text tokens).
We conduct experiments to investigate the importance of core tokens
when we study the effectiveness of the spanning-based technique in the next section.

\subsection{Filtering candidates}
\label{paper:2ED-pruning}

In the Filtering Candidates component of our algorithm (cf. Figure~\ref{fig:framework}),
we can integrate different filtering (i.e. pruning) techniques for the two-level similarity used in the Measuring Similarity component.
Next, we propose a filtering technique for FuzzyED, and present a general filtering technique for both FuzzyED and Fuzzy Jaccard.

\textbf{A filtering technique for FuzzyED}:
The key idea of the filtering technique is to compute a lower bound cost on transforming a sub-string candidate to an entity,
and to prune the sub-string candidate if the lower bound cost is higher than a certain threshold.
The lower bound cost includes the insertion and substitution cost on
transforming the sub-string candidate $\mathcal{S}$ to the entity $\mathcal{E}$,
and is computed by the equation below.
\begin{small}
\begin{equation}
C_\perp(\mathcal{E}, \mathcal{S}) = \sum_{\mathcal{E}_i \in \mathcal{E}}{(1 - M_i) \times w(\mathcal{E}_i)}
\label{eq:pruning}
\end{equation}
\end{small}%
where $M_i$ is the edit similarity of the entity token $\mathcal{E}_i$ to
the most similar text token in the sub-string candidate $\mathcal{S}$;
$M_i \in [0, 1]$.
Note that both the insertion cost and the substitution cost are considered in the above equation,
because $M_i = 0$ is the case of insertion and $M_i > 0$ is the case of substitution.
Note also that we do not include the deletion cost in Equation~\eqref{eq:pruning},
because our algorithm finds the most similar sub-string of $\M{S}$ that matches the entity
and we do not know if the unmatched tokens are part of the most similar sub-string.
If the lower bound cost $C_\perp(\mathcal{E}, \mathcal{S})$ is higher than the threshold $(1 - \delta)$,
we prune the sub-string candidate and avoid measuring the FuzzyED similarity.

\textbf{A general filtering technique}:
In the Filtering Candidates component, we can use more than one filtering technique.
Here, we propose to use one more filtering technique introduced by Chakrabariti et al.~\cite{chakrabarti2008efficient}.
Formally, a sub-string candidate can be pruned if the condition below is satisfied.
\begin{small}
\begin{equation}
T_w(\M{S} \cap \M{E}) < \delta
\end{equation}
\end{small}%
The technique can be used in FuzzyED and Fuzzy Jaccard.
No other proper filtering techniques come to our awareness for Fuzzy Jaccard.
Hence, we use the above filtering technique in our algorithm using the Fuzzy Jaccard similarity.

\section{Experimental Study}
\label{paper:es}
In this section, we present the experimental results on the efficiency and effectiveness
of our algorithm using FuzzyED denoted by ``FED" and our algorithm using Fuzzy Jaccard denoted by ``FJ''.
In the Matching Text Token component of our algorithm,
we used the C++ source code offered by Li et al.~\cite{li2011faerie}.
We implemented FED and FJ in C++.
All experiments were conducted on a machine running Linux with an Intel Xeon E5-2643 CPU and 32GB memory.
By default, we set the entity similarity threshold $\delta$ to 0.9,
and the token edit similarity threshold $\tau$ to 0.8.
We used three real world datasets: Amazon Reviews dataset~\cite{mcauley2013hidden}, DBWorld Messages dataset
and IMDB Reviews dataset~\cite{maas2011learning}.
The details of the datasets are as follows.
(i) Amazon Reviews: the dataset contains 346,867 product reviews from the customers of Amazon.
Each product review serves as a document;
1,989,376 product names from Amazon form the entity dictionary.
(ii) DBWorld Messages: the dataset contains 33, 628 messages of ``call for papers", job advertisement and so forth
in the database research community.
Each message is a document; the entity dictionary contains 132,745 worldwide institution names from Free-base~\cite{bollacker2008freebase}.
(iii) IMDB Reviews: the dataset has 97,788 movie reviews from the IMDB website.
Each movie review is a document; the entity dictionary contains 108,941 movie names in the IMDB website.
More details of the three datasets are provided in Table~\ref{tbl:dataset-info};
the average, maximum and minimum length of the documents (or the entities in the dictionary) are measured by the number of tokens.

\begin{table}
\centering
\begin{small}
\caption{Details of documents and dictionaries}
\begin{tabular}{|c|c|c|c|c|} \hline
dataset			& size 		& ave len	& max len	& min len\\\hline
Amazon doc		& 346,867	& 191		& 29,070		& 30		\\
Amazon dict		& 1,989,376	& 6 			& 204 		& 1 		\\\hline
DBWorld doc		& 33,628		& 732		& 33,648		& 1		\\
DBWorld dict		& 132,745	& 3			& 27			& 1 		\\\hline
IMDB doc			& 97,788		& 277		& 2,968		& 8		\\
IMDB dict		& 108,941	& 3			& 24			& 1		\\\hline
\end{tabular}
\end{small}
\label{tbl:dataset-info}
\TSPACE
\end{table}

We have four implementations of our algorithm:
\textit{FED-e} (\textit{FED-s}) is FuzzyED together with the enumeration-based (spanning-based) candidate producing technique;
\textit{FJ-e} (\textit{FJ-s}) is Fuzzy Jarcard together with the enumeration-based (spanning-based) candidate producing technique.

In what follows, we first report the efficiency and effectiveness of our algorithm,
and then we investigate the effect of core tokens on our algorithm.

\subsection{Efficiency and effectiveness comparison}
Here, we investigate the performance of our algorithm in three aspects: overall efficiency,
the effect of varying the parameters (e.g. $\tau$ and $\delta$) on the efficiency,
and effectiveness.

\eat{
\captionsetup[subfloat]{captionskip=-35pt}
\begin{figure*}
\centering

\subfloat[\vspace{50pt} Varying entity similarity threshold \label{fig:vary-entity-sim}]{
  \includegraphics[width=2.1in, height=2.2in]{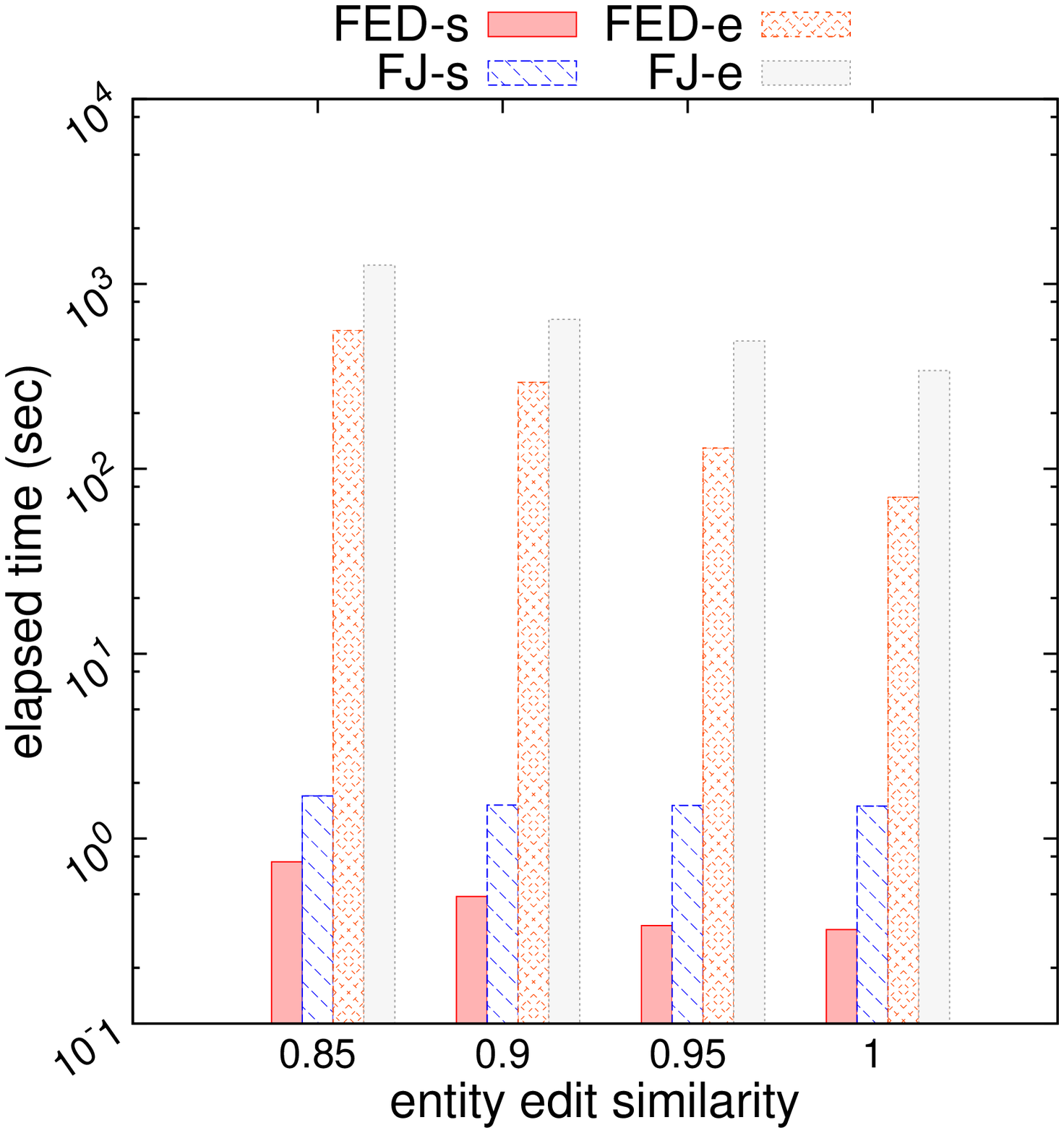}

}
\subfloat[\vspace{50pt} Varying token similarity threshold\label{fig:vary-ed-sim}]{
\includegraphics[width=2.1in, height=2.2in]{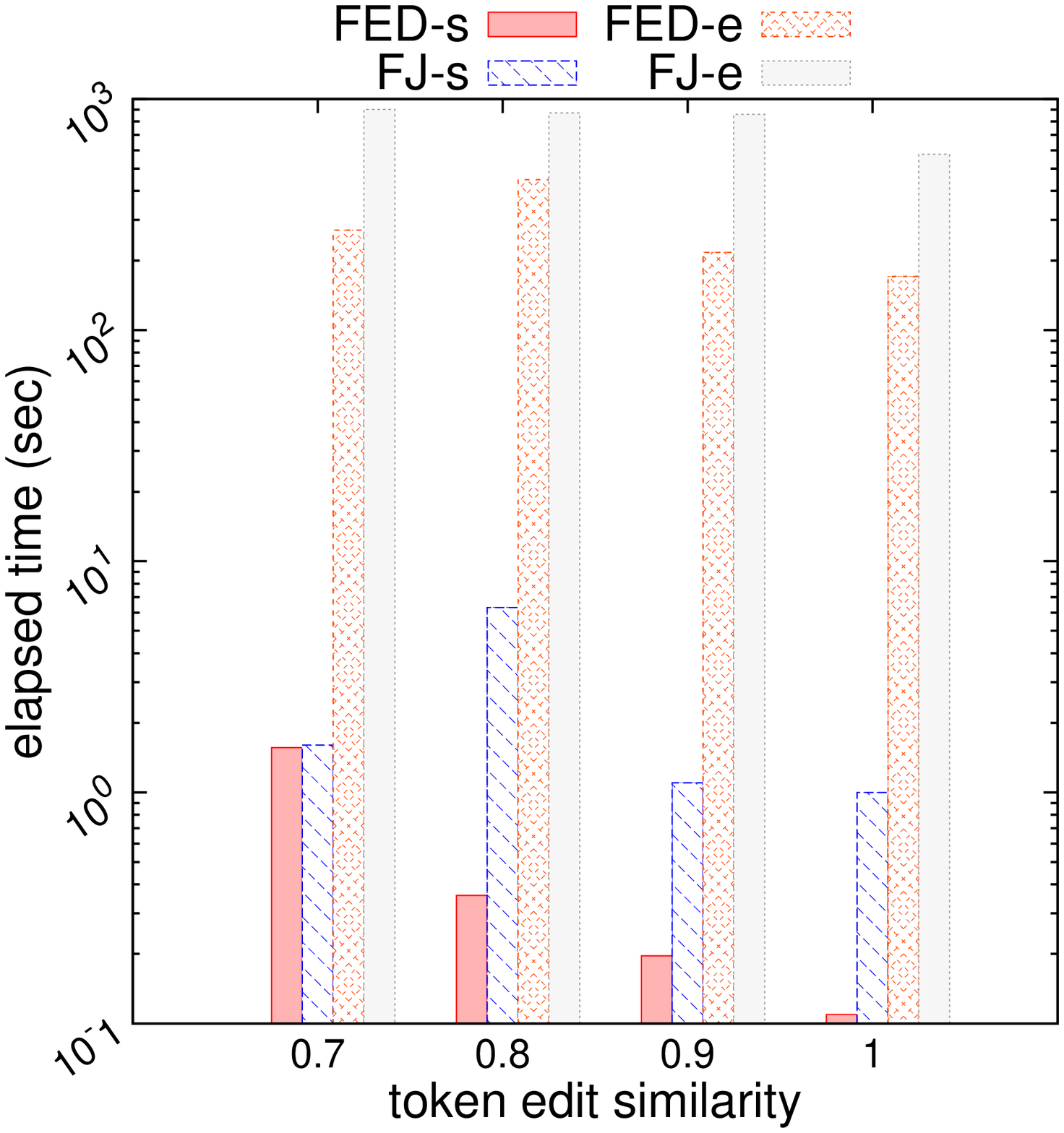}
}
\subfloat[\vspace{50pt} Varying dictionary size \label{fig:vary-dict-size}]{
\includegraphics[width=2.1in, height=2.2in]{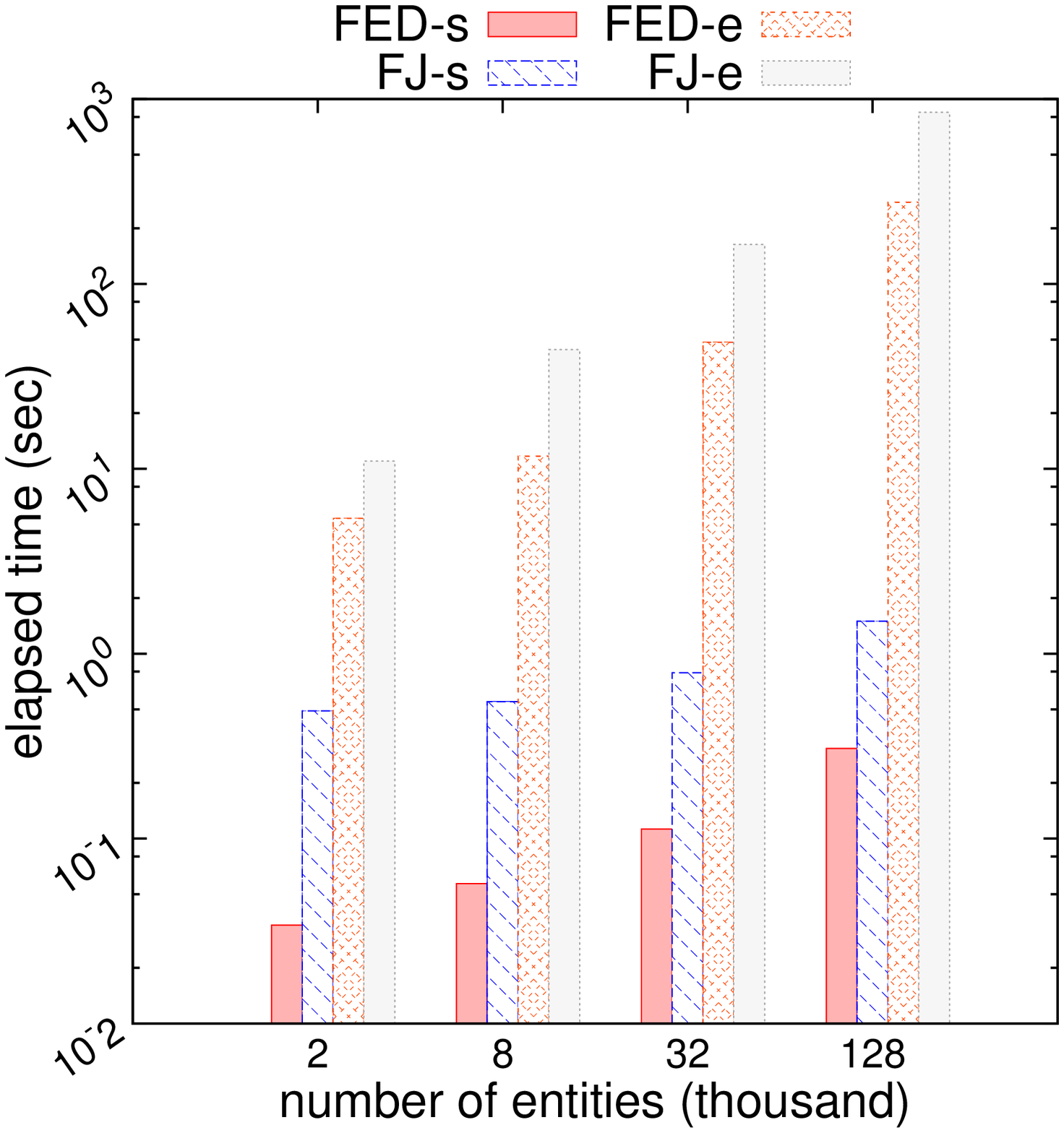}
}
\vspace{-65pt}
\caption{Varying similarity thresholds and the size of the dictionary}
\vspace{-15pt}
\end{figure*}
}

\subsubsection{Overall efficiency}
\label{paper:es:ovall-eff}

We conducted experiments on the three datasets for FED-s
and the elapsed time of FED-s for Amazon Reviews, DBWorld Messages and IMDB Reviews
is 16 hours, 17 minutes and 16 minutes, respectively;
FJ-s took twice more time than FED-s to process the datasets.
Note that the Amazon Reviews dataset has around half a million documents
and two million entities in the dictionary,
our FED-s can process it in 16 hours.
In comparison, FED-e and FJ-e are extremely slow to process the whole datasets,
because they require measuring the two-level similarity for more sub-string candidates as discussed in Section~\ref{paper:2ED-s-analysis}.
To provide some specific results on the elapsed time of the four implementations,
we randomly sampled a sub-dataset from each of the original document dataset.
To construct the three sub-datasets, we sampled 1 per 100 documents in the DBWorld dataset and in the IMDB dataset,
and 1 per 10,000 documents in the Amazon dataset.
Thus, FJ-e and FED-e can process the three sub-datasets in a reasonable amount of time.
Note that we do not construct subsets of the dictionaries
and we show the effect of changing the size of the dictionary in the next set of experiments.

\begin{table}
\centering
\begin{small}
\caption{Overall efficiency comparison}
\begin{tabular}{|c|c|c|c|c|} \hline
\multirow{2}{*}{sub-dataset} & \multicolumn{2}{c|}{using enumeration} & \multicolumn{2}{c|}{using spanning}\\\cline{2-5}
			& FED-e	& FJ-e			& FED-s		& FJ	-s			\\\hline
Amazon 		& 1.05 h		& 26.7 h	 	& 7	sec		& 10 sec		\\
DBWorld		& 0.25 h		& 12.9 h		& 11	 sec		& 11 sec		\\
IMDB			& 0.13 h		& 6.06 h		& 11	 sec		& 12 sec		\\\hline
\end{tabular}
\label{tbl:overall-efficiency}
\end{small}
\TSPACE
\end{table}

Table~\ref{tbl:overall-efficiency} gives the efficiency of the four implementations on the three sub-datasets.
As we see from the table,
implementations using spanning-based candidate producing technique (i.e. FED-s and FJ-s) are more than 40 times
faster than those using enumeration-based candidate producing technique.
Another observation is that FED based implementations are more efficient than FJ based implementations,
because FJ has higher complexity than FED as discussed in Section~\ref{paper:fj-analysis}.

\captionsetup[subfloat]{captionskip=-28pt}
\begin{figure}
\vspace{-5pt}
\centering

\subfloat[\vspace{50pt} Varying $\delta$ \label{fig:vary-entity-sim}]{
  \includegraphics[width=1.6in, height=1.6in]{vary-entity-similarity.eps}

}
\subfloat[\vspace{50pt} Varying $\tau$ \label{fig:vary-ed-sim}]{
\includegraphics[width=1.6in, height=1.6in]{vary-ed-sim.eps}
}
\vspace{-60pt}

\subfloat[\vspace{50pt} Varying dictionary size \label{fig:vary-dict-size}]{
\includegraphics[width=1.6in, height=1.6in]{vary-dict-size.eps}
}
\subfloat[\vspace{50pt} Varying \# of doc. \label{fig:vary-numof-doc}]{
  \includegraphics[width=1.6in, height=1.6in]{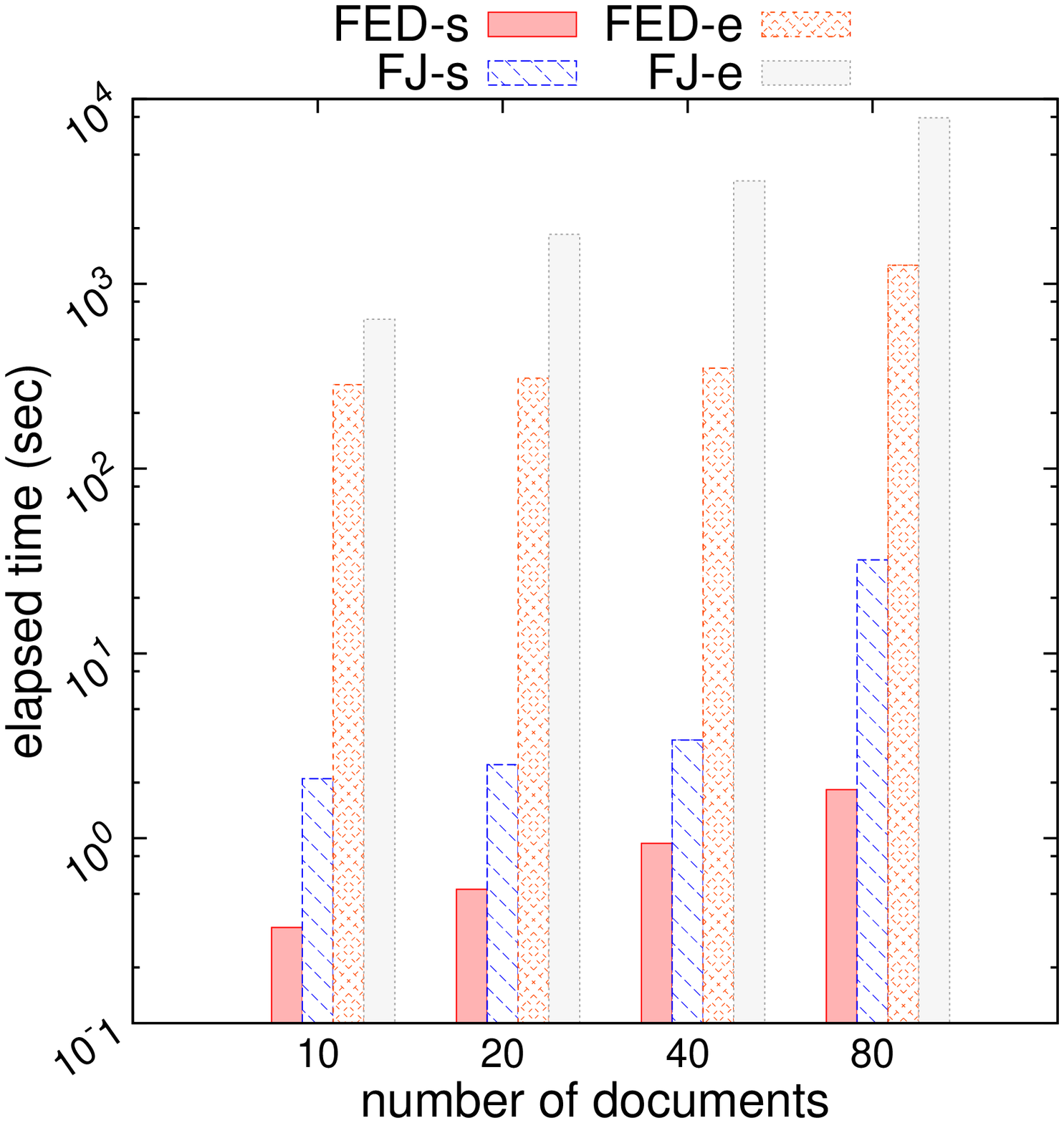}

}
\vspace{-55pt}
\caption{Varying different parameters}
\vspace{-15pt}
\end{figure}

\subsubsection{Effect of varying the parameters on efficiency}
Next, we study the effect of varying the parameters on the efficiency of FED-s, FED-e, FJ-s and FJ-e.
In our experiments, we observed that the results on the three datasets are similar
when varying different parameters.
Due to the space limitation, we use the DBWorld Messages dataset as a representative in this set of experiments.
The default settings of the experiments are as follows:
(i) the entity similarity threshold $\delta$ is set to 0.9; (ii) the token edit similarity threshold $\tau$ is set to 0.8;
(iii) the number of entities in the dictionary is 132,745 (i.e. the whole dictionary) and (iv) the number of documents is 10.

\textbf{Effect of varying the entity similarity threshold}:
To study the effect of the entity similarity threshold $\delta$, we varied $\delta$ from 0.85 to 1.
Figure~\ref{fig:vary-entity-sim} shows the results of the effect on the four implementations.
As can be seen from the figure, FED based implementations consistently outperform FJ based implementations.
Implementations using spanning-based candidate producing technique outperform
those using enumeration-based candidate producing technique by around 100 times.
An observation of the figure is that as the entity similarity threshold decreases
the total elapsed time of all the implementations increases.
This is because when the entity similarity threshold is small,
more candidates require measuring the two-level similarity.

\textbf{Effect of varying the token similarity threshold}:
Figure~\ref{fig:vary-ed-sim} gives the results of varying the token similarity threshold $\tau$ from 0.7 to 1.
FED-s and FJ-s significantly outperform FED-e and FJ-e by two orders of magnitude.
Similar to varying the entity similarity threshold $\delta$,
the smaller the threshold, the more time our algorithm requires.
\eat{
This is because when the threshold is small, more tokens in the document match entities
and hence the number of sub-strings generated increases.}

\textbf{Effect of varying the size of the entity dictionary}:
To study the effect of the size of the dictionary,
we varied the number of entities in the dictionary from 2,000 to 128,000.
Figure~\ref{fig:vary-dict-size} shows that the elapsed time of all the four implementations increases as the size of the dictionary increases.

\eat{
\captionsetup[subfloat]{captionskip=-35pt}
\begin{figure}
\vspace{-5pt}
\centering

\subfloat[\vspace{50pt} Varying \# of doc. \label{fig:vary-numof-doc}]{
  \includegraphics[width=1.7in, height=2in]{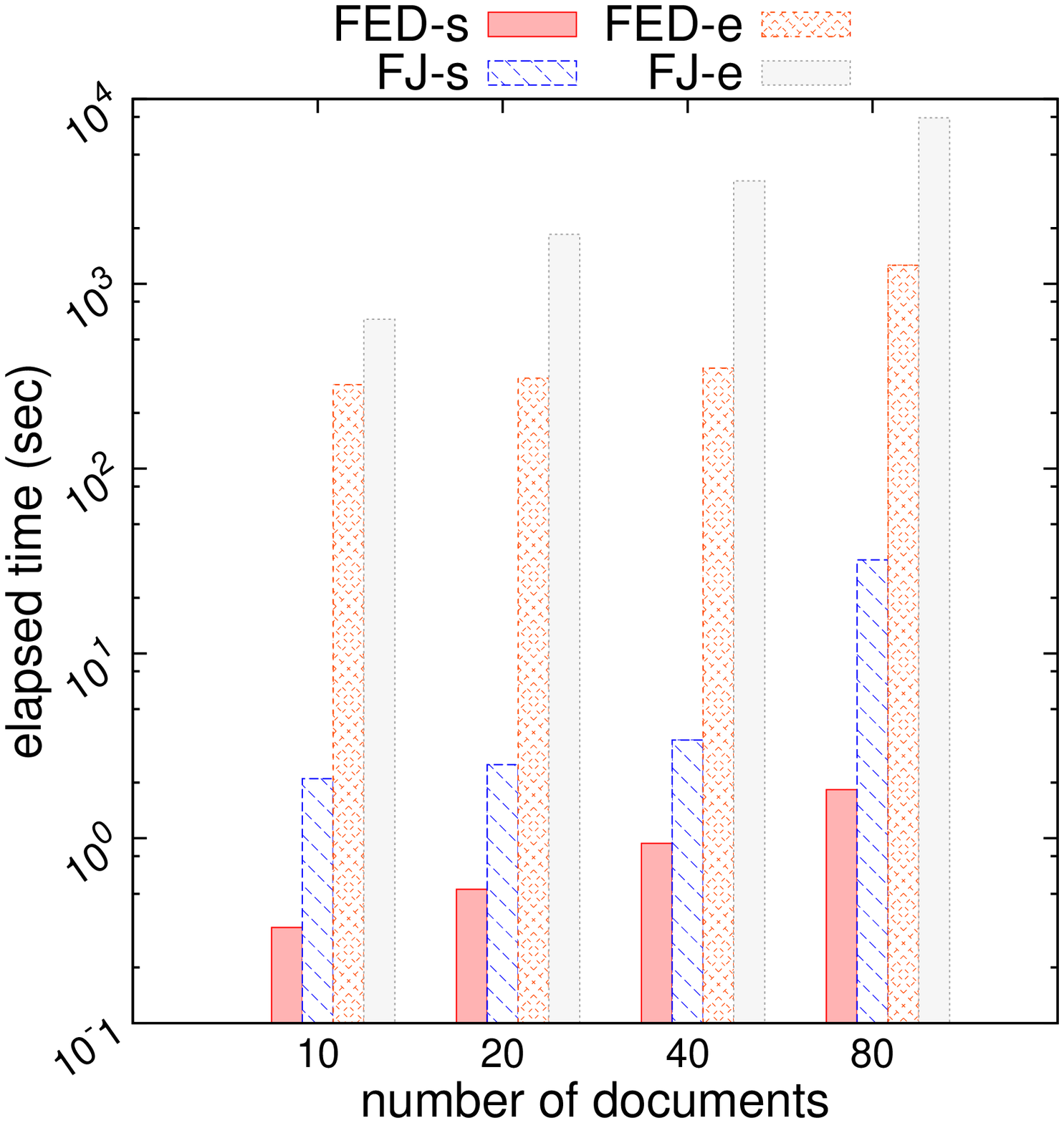}

}
\subfloat[\vspace{50pt} Varying doc. len. \label{fig:vary-lenof-doc}]{
\includegraphics[width=1.7in, height=2in]{vary-lenof-doc.eps}
}
\vspace{-55pt}
\caption{Varying the document dataset}
\end{figure}
}

\textbf{Effect of varying the number of documents}:
To study the effect of the number of documents on the efficiency,
we sampled from the DBWorld Messages dataset four sub-datasets of 10, 20, 40 and 80 documents with the average length of 732.
We measured the total elapsed time of extracting entities from each sub-dataset.
As shown in Figure~\ref{fig:vary-numof-doc},
the elapsed time of FED based implementations grows more slowly compared with FJ based ones.
This is because the more documents, the more sub-string candidates are generated.
As a result, our algorithm needs to measure more two-level similarity.
As the cost on measuring the two-level similarity of FED based implementations is cheaper than that of FJ based ones
($O(mn)$ v.s. $O(m^2n^2)$),
the elapsed time of FJ based implementations increases faster than that of FED based ones.

\subsubsection{Overall effectiveness}

To demonstrate the effectiveness of the FuzzyED similarity and the Fuzzy Jaccard similarity, we used the whole dataset of DBWorld Messages.
We manually labelled 20,000 sub-string candidates as a set of ground truth.
Entities in the document correctly extracted as entities in the dictionary are called true positive (denoted by $tp$);
no entities in the document extracted as entities in the dictionary are called false positive (denoted by $fp$).
We compute the precision $p$ and recall $r$ by the following equations.
\begin{small}
\begin{equation*}
p=\frac{tp}{tp+fp}
\text{ \qquad \ } r=\frac{tp}{tp+fn}
\end{equation*}
\end{small}%
where $fn$ is the number of false negative and hence $tp + fn$ is the total number of true positives in the ground truth set.

\begin{table}
\centering
\begin{small}
\caption{F-measure of FED and FJ}
\begin{tabular}{|c|c|c|c|c|c|c|c|} \hline
\multirow{2}{*}{$\delta$} & \multicolumn{2}{c|}{precision}
& \multicolumn{2}{c|}{recall} & \multicolumn{2}{c|}{F$_1$}\\\cline{2-3} \cline{4-5} \cline{6-7}
				& FED 			& FJ			& FED		& FJ	 	 	& FED		& FJ	 	\\\hline
1.00 			& 100\%			& 97.6\% 	& 94.5\%		& 95.4\%		& 97.2		& 96.5	\\
0.95				& 88.0\%			& 85.3\%		& 94.8\%		& 95.5\%		& 91.3		& 88.4	\\
0.90				& 71.5\%			& 69.5\%		& 96.6\%		& 97.1\%		& 82.2		& 81.0\\
0.85				& 64.1\%			& 62.6\%		& 99.7\%		& 100\%		& 78.0		& 77.0\\\hline
\end{tabular}
\end{small}
\label{tbl:effectiveness}
\TSPACE
\end{table}

Table~\ref{tbl:effectiveness} shows the results of F-measure for FED and FJ
on the entity similarity threshold $\delta$ changing from 0.85 to 1.
As we can see from the table, FED has better F$_1$ score and precision than FJ and comparable recall to FJ.
FED and FJ can produce an F$_1$ score of around 0.9 when the entity similarity threshold is 0.95.
\eat{An observation from the table is that FED can achieve 100\% precision while FJ cannot
even when the entity similarity threshold is set to 1.
This is because FJ does not consider order and has no mechanism to distinguish sub-strings
such as ``University Park" from the entity ``Park University".
}

\subsection{Effect of core tokens}
In this set of experiments, we provide experimental results of the spanning-based approach using core tokens
compared with the spanning-based approach without using core tokens as discussed in Section~\ref{paper:2ED-s-analysis}.
The datasets used in these experiments are identical to those detailed in Table~\ref{tbl:overall-efficiency}.

\begin{table}
\centering
\begin{small}
\caption{Effect of core tokens on candidates producing}
\begin{tabular}{|c|c|c|c|} \hline
sub-dataset	& FED-s		& FED-a	 	& speedup	\\\hline
Amazon 		& 7	sec	& 0.70 hr 		& 362	\\
DBWorld		& 11	 sec	& 0.14 hr 		& 45		\\
IMDB			& 11	 sec	& 0.17 hr		& 56	\\\hline
\end{tabular}
\end{small}
\label{tbl:eff-seg-core}
\TSPACE
\vspace{-10pt}
\end{table}

To demonstrate the effectiveness of using core tokens,
we used two versions of FED: one with core tokens applied in the candidate producing process;
the other, denoted by FED-a (``a'' for all entity tokens), without using core tokens in the candidate producing process.
Note that the FJ based approach without using core tokens are extremely slow and did not complete within our time limit,
and hence the results of FJ are not shown here.
As we can see from Table~\ref{tbl:eff-seg-core}, FED-s consistently outperforms FED-a by upto 362 times.
This is because using core tokens reduces the number of matched tokens in the document,
and hence significantly reduces the number of sub-string candidates which requires measuring the two-level similarity.

\section{Conclusion}
\label{paper:con}
In this paper, we have addressed the problem of entity extraction from free text using
both character-based similarity and token-based similarity (i.e. two-level similarity).
By exploiting the properties of the two-level similarity and the weights of tokens,
we have developed novel techniques to significantly
reduce the number of sub-string candidates that require computation of two-level similarity against the entities.
A comprehensive experimental study has shown that
our algorithm based on edit similarity is efficient and also effective.
Moreover, our algorithm produces a high F$_1$ score in the range of [0.91,0.97] with edit similarity of [0.95,1].

\begin{small}
\bibliographystyle{IEEEtran}
\bibliography{FuzzyED}
\end{small}

\appendix

\section{Proof of Lemma 1}
\label{appendix:core-proof}

We prove Lemma~\ref{lemma-core} for FuzzyED in the following.
\begin{proof}
Suppose no token in $\mathcal{S}$ matches the core tokens of $\mathcal{E}$.
For transforming $\mathcal{S}$ to $\mathcal{E}$,
at least we need to insert all the core tokens in $\mathcal{E}$ to $\mathcal{S}$
and the total cost of the insertion is larger than $1 - \delta$ (cf. Constraint~\eqref{eq:core-ed}).
Hence, the similarity between $\mathcal{S}$ and $\mathcal{E}$ is smaller than $\delta$.
Therefore, for $\mathcal{S}$ to match $\mathcal{E}$ (i.e. the similarity between $\mathcal{S}$ and $\mathcal{E}$ is not smaller than $\delta$),
at least one text token in $\mathcal{S}$ must match a core token of the entity $\mathcal{E}$.
\end{proof}
Similarly, we can prove Lemma~\ref{lemma-core} for Fuzzy Jaccard.

\section{Proof of Lemma 2}
\label{appendix:fj-bound}

We prove Lemma~\ref{lemma-increase} for the Fuzzy Jaccard similarity here.
\begin{proof}
Only the token that improves the Fuzzy Jaccard similarity are included in the current sub-string.
In what follows, we investigate tokens that improve the Fuzzy Jaccard similarity.
Since $\M{E}_r$ and $\M{S}'_r$ are similar and we assume they have the same IDF value,
we replace $\M{E}_r$ by $\M{S}'_r$ in the following process.
As $\M{S}'_r$ leads to increase of the Fuzzy Jaccard similarity, we have
\begin{small}
\begin{equation*}
\frac{\sum{M_i\cdot idf(\M{S}'_i)} + (1-M_r)idf(\M{S}'_r)}{T_{idf}(\M{S}) + idf(\M{S}'_r)} \geq
\frac{\sum{M_i\cdot idf(\M{S}'_i)}}{T_{idf}(\M{S})}.
\end{equation*}
\end{small}%
We let $a=\sum{M_i\cdot idf(\M{S}'_i)}$, $b=T_{idf}(\M{S})$, $c=idf(\M{S}'_r)$, $c'=(1-M_r) \cdot idf(\M{S}'_r)$.
Then we have
\begin{small}
\begin{equation*}
\frac{a + c'}{b + c} \geq \frac{a}{b}.
\end{equation*}
\end{small}%
Since $a$, $b$, $c$ and $c'$ are larger than 0, we can rewrite the above inequality as follows.
\begin{small}
\begin{align*}
 ab + c'b \geq ab + ac  \hspace{10pt}
\Rightarrow  \hspace{0pt} c'b \geq ac \hspace{10pt}
\Rightarrow  \hspace{0pt} \frac{c'}{c} \geq \frac{a}{b}
\end{align*}
\end{small}%
Substituting the original values of $a$, $b$, $c$ and $c'$, we have
\begin{small}
\begin{equation*}
(1-M_r) \geq \frac{\sum{M_i\cdot idf(\M{S}'_i)}}{T_{idf}(\M{S})}.
\end{equation*}
\end{small}
\end{proof}
Only tokens that satisfy the above constraint are included in the spanning process for Fuzzy Jaccard.

\end{document}